\documentclass{bmcart}

\usepackage[utf8]{inputenc} 

\usepackage{booktabs} 

\usepackage{subfig}
\usepackage{multicol}
\usepackage{amsmath}
\usepackage{amssymb}
\usepackage{bbm}
\usepackage[contents={}]{background}
\usepackage{color}

\usepackage{footnote}
\usepackage{url}
\usepackage{epsfig}

\def\includegraphics{}

\startlocaldefs
\endlocaldefs

\begin{document}

\begin{frontmatter}

\begin{fmbox}
\dochead{Research}


\title{Mining large-scale human mobility data for long-term crime prediction}


\author[
   addressref={aff1},                   
   email={ckadar@ethz.ch}   
]{\inits{CK}\fnm{Cristina} \snm{Kadar}}
\author[
   addressref={aff1},
   email={irena.pletikosa@gmail.com}
]{\inits{IP}\fnm{Irena} \snm{Pletikosa}}


\address[id=aff1]{                        
  \orgname{ETH Zurich, D-MTEC}, 
  \street{Weinbergstrasse 56/58},                  
  \postcode{8092}                                
  \city{Zurich},                             
  \cny{Switzerland}                                    
}


\begin{artnotes}
\end{artnotes}

\end{fmbox}


\begin{abstractbox}

\begin{abstract} 
Traditional crime prediction models based on census data are limited, as they fail to capture the complexity and dynamics of human activity. With the rise of ubiquitous computing, there is the opportunity to improve such models with data that make for better proxies of human presence in cities. In this paper, we leverage large human mobility data to craft an extensive set of features for crime prediction, as informed by theories in criminology and urban studies. We employ averaging and boosting ensemble techniques from machine learning, to investigate their power in predicting yearly counts for different types of crimes occurring in New York City at census tract level. Our study shows that spatial and spatio-temporal features derived from Foursquare venues and checkins, subway rides, and taxi rides, improve the baseline models relying on census and POI data. The proposed models achieve absolute $R^2$ metrics of up to $65\%$ (on a geographical out-of-sample test set) and up to $89\%$ (on a temporal out-of-sample test set). This proves that, next to the residential population of an area, the ambient population there is strongly predictive of the area's crime levels. We deep-dive into the main crime categories, and find that the predictive gain of the human dynamics features varies across crime types: such features bring the biggest boost in case of grand larcenies, whereas assaults are already well predicted by the census features. Furthermore, we identify and discuss top predictive features for the main crime categories. These results offer valuable insights for those responsible for urban policy or law enforcement.
\end{abstract}


	

\end{abstractbox}
%

\end{frontmatter}



\section{Introduction}
 
Crime prediction is inherently difficult. Crime analysis has already confirmed that crimes are unequally distributed in time and space \cite{Brantingham1993}. Furthermore, crime is a highly dynamic and complex phenomenon driven by the people and the environment where they meet \cite{Miethe1994}, and scholars in different disciplines are still investigating various elements for predictive power. Knowing when and where crime is more likely to occur can help various actors engaged in crime reduction: urban planners to design safer cities \cite{Clarke2009} and police forces to better direct their patrols \cite{Braga2005}.


Initially, criminological studies have focused solely on socio-demographic attributes as factors correlating with victimization and have noticed that specific groups of people tend to have lifestyles that exposed them to higher risk of victimization compared to other groups --  as explained by the \textit{Lifestyle Exposure Theory} \cite{Hindelang1978}. For instance: men, young adults, and African Americans have been found to experience higher risk of victimization in general \cite{Hindelang1978}. Under the umbrella of the \textit{Social Disorganization Theory}, a series of criminological studies have explained crime as a product of the ecological attributes of the neighborhood: ethnicity, income level, and residential stability \cite{Pratt2005, Sampson1997}. 

Cohen and Felson extended the model beyond the attributes of the underlaying populations towards opportunity -- according to their \textit{Routine Activity Theory} \cite{Cohen1979} there are three elements which need to be present in time and space for a crime to occur: a motivated offender, a suitable target, and a lack of guardianship. 
Finally, Brantingham and Brantingham analyzed criminogenic places in cities -- places that make crime easy and profitable and are the by-products of the environments we build to support the requirements of everyday life (e.g. homes, shops, offices, government buildings, parks, bus stops or sports stadia) \cite{Brantingham1995} -- and divided them into crime attractors and crime generators. Crime attractors are places which attract criminals, because there are known opportunities in those areas. As a consequence, the probability of a crime happening in those places is higher compared to other places (e.g. night life district). In turn, crime generators are places in which crime emerges at times where large number of people are attracted to those places for reasons other than to offend (e.g. massive sports events).\footnote{We have limited our survey of theories in criminology to the main theories that look at victims and offenders and their routine activities, and are relevant for this study. Indeed, there are also other factors that influence criminal behavior, such the attributes of the built environment. For instance, Wilson and Kelling proposed in their \textit{Broken Windows Theory} \cite{Wilson1982} that degraded urban environments (such as broken windows, graffiti, excessive litter) enhance criminal activities in the area.}

Other, more qualitative works in urban planning, have also looked at the relationship between the built environment, population and safety. Specifically, two notable works do not agree whether the density and diversity of human activity within an area are attracting crime or not. In the \textit{Eyes on the Street Theory} \cite{Jacobs1961}, Jacobs postulates that higher densities of people and buildings, pedestrian areas and a mix of activities in the neighborhood act as crime deterrents. On the other hand, Newman suggests that less built areas with more segregated activities are safer \cite{Newman1973}. 

In terms of data, traditionally, quantitative models explaining crime have leveraged the socio-demographic and economical data available from the census, describing the resident population of a given neighborhood \cite{Hindelang1978,Sampson1997}. From a theoretical point of view, these models have relied on the initial victimization theories in criminology.

But census data has an intrinsic limitation, in that it only offers a static and sometimes obsolete image of the city, without capturing the people dynamics over time and space. There is now the opportunity for non-conventional factors to be integrated in crime prediction models by tapping into novel data sources that reflect the structure and dynamics of our cities. With the emergence of mobile phones and other types of ubiquitous computing, a plethora of geo-tagged crowd-generated data can now offer an approximation of the ambient population. In particular, location-based social networks (LBSNs) like Foursquare offer a very vivid image of the city, being able to not only provide time and location of human activity, but also the context (like traveling, shopping, working, going out, etc.) in which activities occur. For example, researchers have successfully showed that Foursquare can be used to automatically infer urban clusters which reflect the local dynamics and character of life of the area \cite{Cranshaw2012}. Furthermore, mobility data, such as public transportation or taxi data, have the capability of capturing the population in and outer flows in different parts of the city. For example, researchers have mined subway usage data to identify deprived areas in the city \cite{Smith2013}. All this leads to the current unique chance of empirically measuring aspects of criminological theories relying on dynamic data which was previously prohibited at large scale.

Hence, in this work, we investigate the potential of geo-tagged human dynamics data for long-term crime prediction models. We use such data to model crime attractors, crime generators and the ambient population in a neighborhood and add these factors on top of the classical factors from census that model the resident population in a neighborhood. The full models for the total number crime incidents achieve absolute $R^2$ metrics of up to $65\%$ when testing on neighborhoods of the same city which have not been used during the training phase of the models, and up to $89\%$ when testing on the full data of the next year. In comparison to the census-only baselines, this translates to improvements of 30 percentage points (on a geographical out-of-sample test set) and of 7 percentage points (on a temporal out-of-sample test set). Furthermore, we look at the major crime types and show that we can achieve improvements of up to 43 percentage points and of up to 9 percentage points, respectively (for the case of grand larcenies).
 
\section{Related Work}
\label{sec:related}

\subsection{Urban Computing}
Nowadays, sensing technologies and large-scale computing infrastructures produce a variety of big data in urban spaces: geographical data, human mobility, traffic patterns, communication patterns, air quality, etc. The vision of urban computing, an emerging field coined by Zheng and collaborators \cite{Zheng2014}, is to unlock the power of big and heterogeneous data collected in urban spaces and apply it to solve major issues our cities face today. They identify seven application areas of urban computing: urban planning, transportation systems, environmental issues, energy consumption, social applications, commercial applications, and public safety and security. 

A special category of this urban data consists of human dynamics data and researchers in the different application areas started to leverage it. For example, within the urban planning and transportation domains, the authors in \cite{Yuan2012} attempt to infer the functions of different regions in the city of Beijing by analyzing the spatial distribution of commercial activities and GPS taxi traces, while the authors in \cite{Chen2015} mine different urban open data sources including LBSNs in the cities of Washington, D.C. and Hangzhou for optimal bike sharing station placement. Furthermore, for commercial purposes, researchers mine LBSNs for optimal retail store placement \cite{Karamshuk2013} or the London metro data for insights into the financial spending of transport users \cite{Lathia2011}, and a variety of urban big data sources for predicting commercial activeness \cite{Yang2017}. Within the public safety and security sector, scholars have just recently started to investigate the potential use of social media \cite{Gerber2014}, of mobile data \cite{Bogomolov2014}, and of taxi flow data \cite{Wang2016} for the purpose of crime inference/prediction. In a related literature stream, authors in \cite{Venerandi2015} exploit POIs from different sources to build classifiers of urban deprivation (a composite score of seven domains, with crime being just one of them) for neighborhoods in the UK, while authors in \cite{Smith2013}, assess the potential of subway flow data to identify areas of high urban deprivation in the city.

\subsection{Crime Prediction}
Researchers in a wide range of fields like criminology, physics and data mining have looked at predicting crime at various scales and using different techniques. In this section we present a short overview of the existing literature.

One approach is to model crime and cities as complex systems, through the lenses of \textbf{urban scaling laws}. A series of papers has found that crime indicators scale super-linearly with the population sizes of cities \cite{Bettencourt2007, Bettencourt2010, Alves2013, Alves2015}. In general, these studies carry out uni-variate \cite{Alves2013, Alves2015} or multi-variate \cite{Alves2017} analysis of crime, i.e. crime as a function of population or of other socio-economic variables, and at a high aggregation level (that of cities). Also, at lower resolution, researchers have confirmed that crime concentrates regardless of city \cite{Oliveira2017} and have found relevant allometric relations between peace disturbance and the resident population, as well as between property crimes and the floating population \cite{Caminha2017}.
   
At intra-city level and using methods from statistical learning, we distinguish between two types of prediction models. The first type of models, consisting of \textbf{long-term crime prediction models}, aim at modeling long-term crime level by looking at aggregated crime rates over 1 to 5 years. In terms of techniques, these models rely on classical inference models like the, sometimes geographically-weighted \cite{Taylor2015, Wang2016}, Poisson \cite{Osgood2000, Kadar2015} and Negative Binomial \cite{Wang2016} regressions, where the task is to predict crime levels and the performance of the model is evaluated in terms of in-sample goodness of fit.
In terms of data, the traditional models in criminology make use of the classical demographic crime correlates, such as residential instability, ethnic heterogeneity, poverty rates, or income rates \cite{Osgood2000, Taylor2015}. Moving to the data mining community, authors in \cite{Kadar2015} use census data, OpenstreetMap POI data, and features of the road network to predict annual burglary levels for municipalities in Switzerland by means of regularized linear regressions tested on a one year left-out sample. Most recent work on long-term crime prediction \cite{Wang2016} makes use of novel nodal features (Foursquare POI data next to demographic data) and edge features (geographical influence of direct neighbors or as computed by taxi flow data) to explain crime rates at community level by means of geographical linear and negative-binomial regressions. 
Similarly, authors in \cite{Kadar2017} employ spatial econometrics techniques where they compare and contrast the explanatory power of a limited set of census and Foursquare features for aggregated census tract crime levels.

The next category of models is the category of \textbf{short-term crime prediction models}, also called spatio-temporal prediction models, where the dependent variable is aggregated over short time periods varying from 1 day to 1 month. 
The most basic and widely applied model for that is the hot spot model \cite{Eck2005}. It clusters past incidents into regions of high risk (the so-called hot spots) using statistical methods like kernel density estimation (KDE) or mixture models. In this case the past is prologue for the future: crime is likely to occur where crime has already occurred! Another set of models that use crime data only are repeat and near-repeat models. Here, researchers have characterized each location by a dynamic attractiveness variable and have represented each criminal as a random walker \cite{Short2008}, or have adapted self-exciting point processes that were initially developed for earthquake modeling to crime modeling \cite{Mohler2011, DOrsogna2015}. The assumption is that some future crimes will occur very near to current crimes in time and place. The biggest disadvantage of models exploiting solely the historical crime records is that they cannot be generalized to areas without historical data.
The spatio-temporal generalized additive model (ST-GAM) \cite{Wang2011} and the local spatio-temporal generalized additive model (LST-GAM) \cite{Wang2012} start looking at socio-demographic data (like population density, unemployment rate, education level, net income, social aid, etc.), and spatial data (like spatial proximity to bus stations, governmental buildings, pawn shops, night life establishments, stores, parks, etc.), and temporal data (like time of day/week/year, temporal proximity to special events such as football games, etc.) describing a criminal incident. These models are extensions of regression models on grids, where the features can be indexed by time. 
Only very recent research has started to utilize human dynamics data in short-term crime prediction models. Gerber \cite{Gerber2014} has shown that combining topics derived from the Twitter stream with the historical crime density delivered by a standard KDE under a logistic regression model leads to an increase in the prediction performance of hotspots next day versus the standard KDE approach for most of the tested crime types. Combining for the first time demographic data and aggregated and anonymized human behavioral data derived from mobile data, Bogomolov and colleagues were able to obtain an accuracy of almost 70\% when predicting whether a specific area in the city will be a crime hotspot or not within the next day \cite{Bogomolov2014}. 

\section{Research Gap and Contributions}
Our work lies within the category of long-term crime prediction models. Compared to previous work in this literature stream, we make following contributions:
\begin{enumerate}
\item in terms of data, we are the first to craft a comprehensive set of spatial and spatio-temporal features describing the dynamics of human activity in an area, as captured by the usage of social networks, public transportation, and road transportation and use this data describing the ambient population to enhance the traditional set of features describing the resident population as modeled by the census statistics. 
\item in terms of techniques, we employ latest averaging and boosting ensemble techniques from machine learning, which in comparison to the current linear models in literature, can deal with the large number of features described above.
\item in terms of evaluation, we test the models on geographical and temporal out-of-sample test sets, to prove generalization and compare them against a weak-baseline based solely on census data and a against a strong-baseline based on census and POI data. We furthermore compare the individual predictive power of the considered data sources of human mobility: Foursquare venues/checkins, NYC subway rides, and NYC yellow and green taxis rides.
\item in terms of unit of analysis, we analyze crime at a granular level, with counts of various types of urban crime being effectively predicted at a high degree of geographic resolution, namely census tracts. We notice different degrees of predictive performance across the different crime types.
\item in terms of interpretability and unlike most studies within the urban computing community, we motivate the choice of features in criminal theory and discuss and interpret the results of the models in this context.
\end{enumerate}

\section{Datasets}
\label{sec:dataset}
New York City (NYC) is a city that has experienced crime across time, though the levels have dropped since the 1990s \cite{Langan2003}, some attributing the success to new policing tactics and the end of the crack epidemic \cite{Blumstein2000}. Furthermore, as part of an initiative to improve the accessibility, transparency, and accountability of the city government, the NYC Open Data platform\footnote{\url{https://nycopendata.socrata.com/}} provides massive data in machine-readable formats on buildings, streets, infrastructure, businesses, permits, licenses, crime, 311 complaints, public transportation, and many more. Furthermore, NYC's 8.5 million inhabitants leave rich digital footprints of their daily activity in various location-based online services, NYC being the most popular city on Foursquare\footnote{\url{http://www.foursquare.com/}} with about 132 million checkins as of May 2016\footnote{\url{http://www.4sqstat.com/}}. 

\subsection{Crime Data}
\label{section:crime}
The raw crime dataset was downloaded from the NYC Open Data platform. For anonymization reasons, in case the offense has not occurred at an intersection, the New York Police Department (NYPD) projects the location of the incident to the center of the block (street segment). Furthermore, crime complaints which involve multiple offenses are classified according to the most serious offense\footnote{\url{https://data.cityofnewyork.us/Public-Safety/NYPD-7-Major-Felony-Incidents/hyij-8hr7}}. Next to the total number of incidents, we concentrate on the following five felony types: grand larceny (which is the theft of another's property, including money, over a certain value), robbery, burglary, felony assault, and grand larceny of motor vehicle -- leaving out the murder and rape cases which have very different underlying causal mechanisms and are also reported on a higher aggregation level. We keep for analysis the data of the last 2 complete years (2014 and 2015). This yields a total number of 174,682 incidents across the five boroughs of NYC: Bronx, Brooklyn, Manhattan, Queens and Staten Island.

\subsection{Census Data}
The census data for NYC was obtained from two separate sources, the 2010 Decennial Census, as well as the 2010-2014 and the 2011-2015 American Community Survey (ACS). In both cases, the data was fetched from the FTP sites of the US Census Bureau\footnote{\url{http://www.census.gov/}}, and was filtered out to keep only the data on a census tract level.

The Decennial Census includes basic demographic figures, which are based on actual counts of persons dwelling in the US and is conducted only once every 10 years. The Summary File 1, used for this study, includes items describing the population, such as gender, age, race, origin, household relationship, household type and size, family type and size, etc. In addition, housing characteristics are captured through the occupancy/vacancy status and tenure.
The ACS estimates are based on yearly collected survey data over a sample of the US population. For the purposes of this study, the 5-year estimates were used, as the largest and most reliable sample, where the data is available on a census tract (and smaller) geography level. Apart from the demographics, ACS contains a rich set of social, housing and economic features, with residential stability, poverty and income being of interest for this study.

\subsection{Foursquare Venues Data}
The Foursquare dataset was collected via the Foursquare API, using the venues search and venue details endpoints. The Foursquare API has been serving both the Foursquare 8.0 and the Swarm apps since the 2014 split of the original Foursquare app.
While Foursquare continues to provide a local search-and-discovery service for places near a user's current location, Swarm lets the user share their location with friends at different precision levels (at city and neighborhood levels, or by checking-in to a specific venue).

The collected data consists of NYC venues with compact metadata like id, name, location, checkins count (total checkins ever done in that venue), users count (total users who have ever checked in), associated categories, menu, opening and popular hours, user-generated tips, etc. We have queried the API by searching for venues in the proximity of every incident location described previously, and this resulted into an extensive database of 273,149 different venues, that have experienced in total over 122 million checkins since their creation on the platform until the time of the data collection (June 2016). 
From these, 250,926 venues have an assigned category. The Foursquare categories span a broad ontology, headed by the following top ten categories: 
Arts and Entertainment (11,794 venues),
College and University (7,082),
Event (84),
Food (47,590),
Nightlife Spot (11,140),
Outdoors and Recreation (18,011),
Professional and Other Places (64,055),
Residence (14,632),
Shop and Service (62,627),
Travel and Transport (13,911).
The distribution of the top categories across the venues is uneven and biased towards establishments where people go out for services, working, shopping, or dining.


\subsection{Subway Usage Data}
Subway usage data, commonly referred to as turnstile data, is released regularly by the Metropolitan Transportation Authority (MTA) and contains entries and exits audit data, generated from the Control Areas from its three main divisions: Interborough Rapid Transit Company (IRT), Independent Subway System (IND) and Brooklyn-Manhattan Transit Company (BMT). While the original dataset contains data from several other associated agencies, for consistency reasons these were left out of the final dataset since the corresponding stations were not located within NYC, or represent train, bus or cable car stations. We downloaded the turnstile data from the New York State Open Data portal\footnote{\url{https://data.ny.gov/en/browse?q=turnstile}} and the MTA website\footnote{\url{http://web.mta.info/developers/turnstile.html}} for the two full years of 2014 and 2015. In addition, a geocoded list of MTA stations was also obtained from the same portal\footnote{\url{https://data.ny.gov/Transportation/NYC-Transit-Subway-Entrance-And-Exit-Data/i9wp-a4ja}}.

To perform the preliminary data cleaning and combine the two data sources, a careful manual examination of station names was conducted. The goal was to resolve situations where the same station appeared with different names in the turnstile dataset, e.g. both '18 AV' and '18 AVE' where coded as '18 AV', and to unify the names used in both datasets. Once the data was cleaned and merged, each station was further examined for location accuracy, by comparing and adjusting it with the corresponding station geolocation provided by Google Maps.
In the end, 455 distinct subway station locations were compiled. In the two years of analysis, they have experienced almost 21 million turnstile updates (the turnstile counters updated every 4 hours).

\subsection{Taxi Usage Data}
The taxi dataset was downloaded from the official website of the City of New York, specifically the Taxi and Limousine Commission\footnote{\url{http://www.nyc.gov/html/tlc/html/about/trip_record_data.shtml}} and combines the 2014 and 2015 complete records of both yellow and green taxi trips. These are the two types of services permitted to pick up passengers via street hails, thus offering a great footprint of human activity. Furthermore, yellow cabs are concentrated around Manhattan and the two main airports (JFK International Airport and LaGuardia Airport), while green cabs are allowed above the 110th Street in Manhattan and in the outer-boroughs of New York City. With the two datasets joined, we obtain a good coverage of the whole city. The trip records include fields capturing pick-up and drop-off timestamps and locations, next to other meta-data like driver-reported passenger counts and trip distances. We have processed in total over 340 millions taxi drives for this work.

\section{Model Specification}
\label{sec:prediction}

\subsection{Unit of Analysis}
We cast the problem as a regression task on the log-transformed crime counts in each census tract. For each census tract, we sum all crime incidents (total and per crime type) occurring in 2014 and in 2015 within the census tract. We opt for crime counts and not crime rates (which are crime counts normalized by the census population), as we like to show the explicit effect of both the resident population (as recorded by census) and of the ambient population (as recorded by the different proxies) on the raw counts.
As a technical remark: we look in the following at points situated in the area of each census track, buffered by 50 feet (which is half the width of the main Manhattan avenues), to account for potential precision inaccuracies in the different spatial data types and to integrate the crime locations that lie on the bordering streets. The same applies for venues, subway, and pickup/drop-off locations.

\begin{figure} [t!]
\centering
\psfig{file=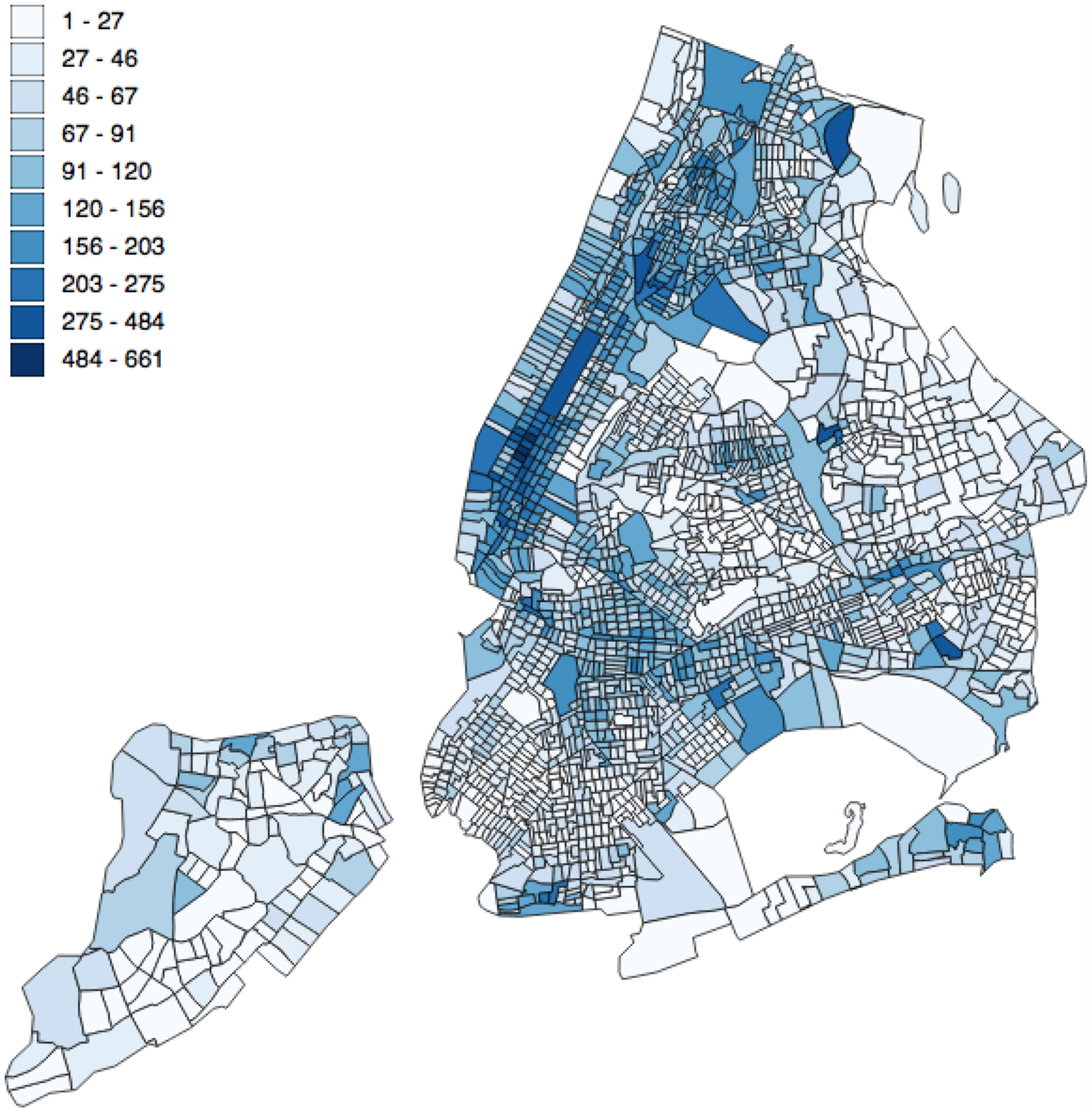, width=0.25\columnwidth}
\hskip 15pt
\psfig{file=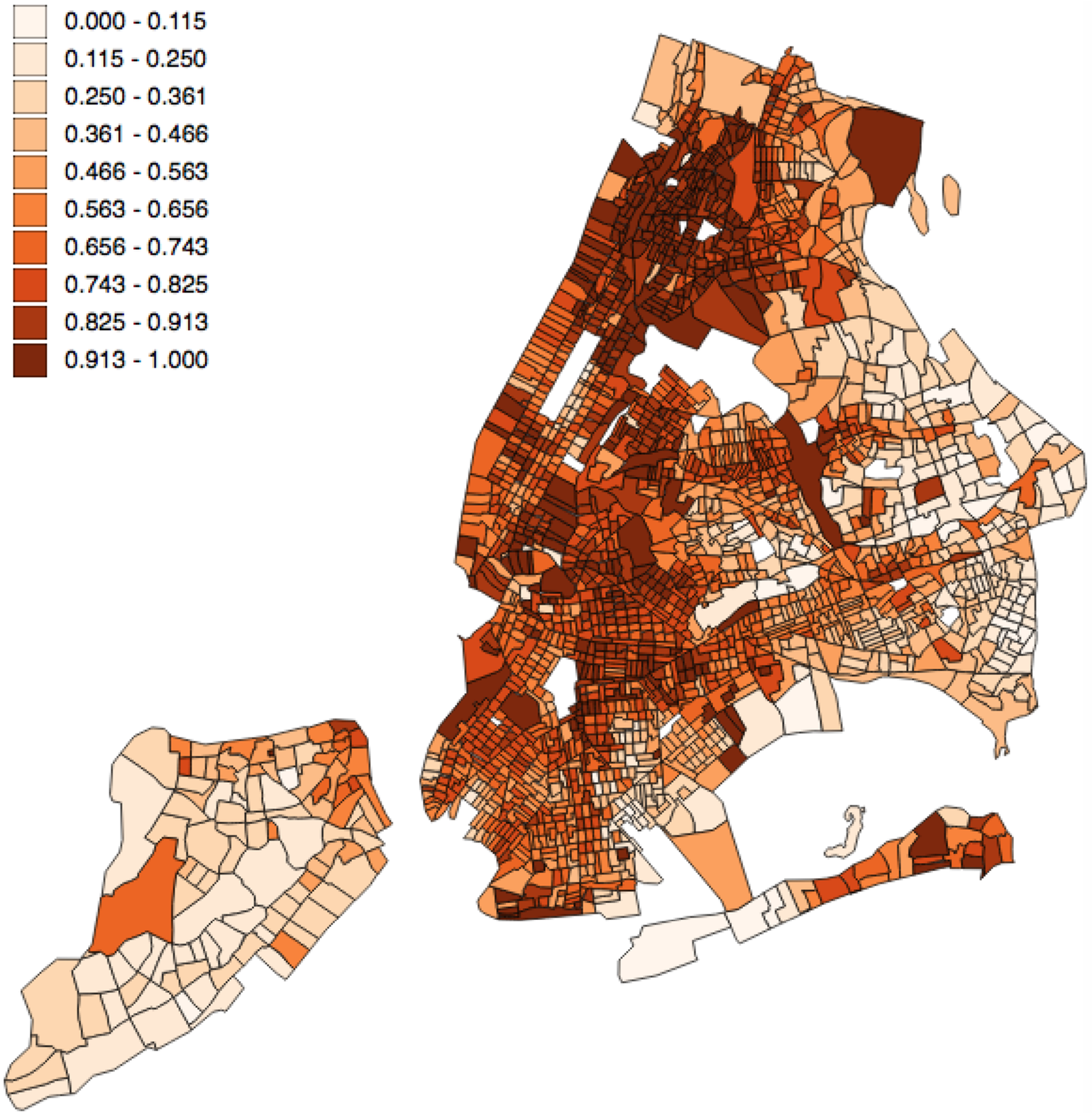, width=0.25\columnwidth}
\hskip 15pt
\psfig{file=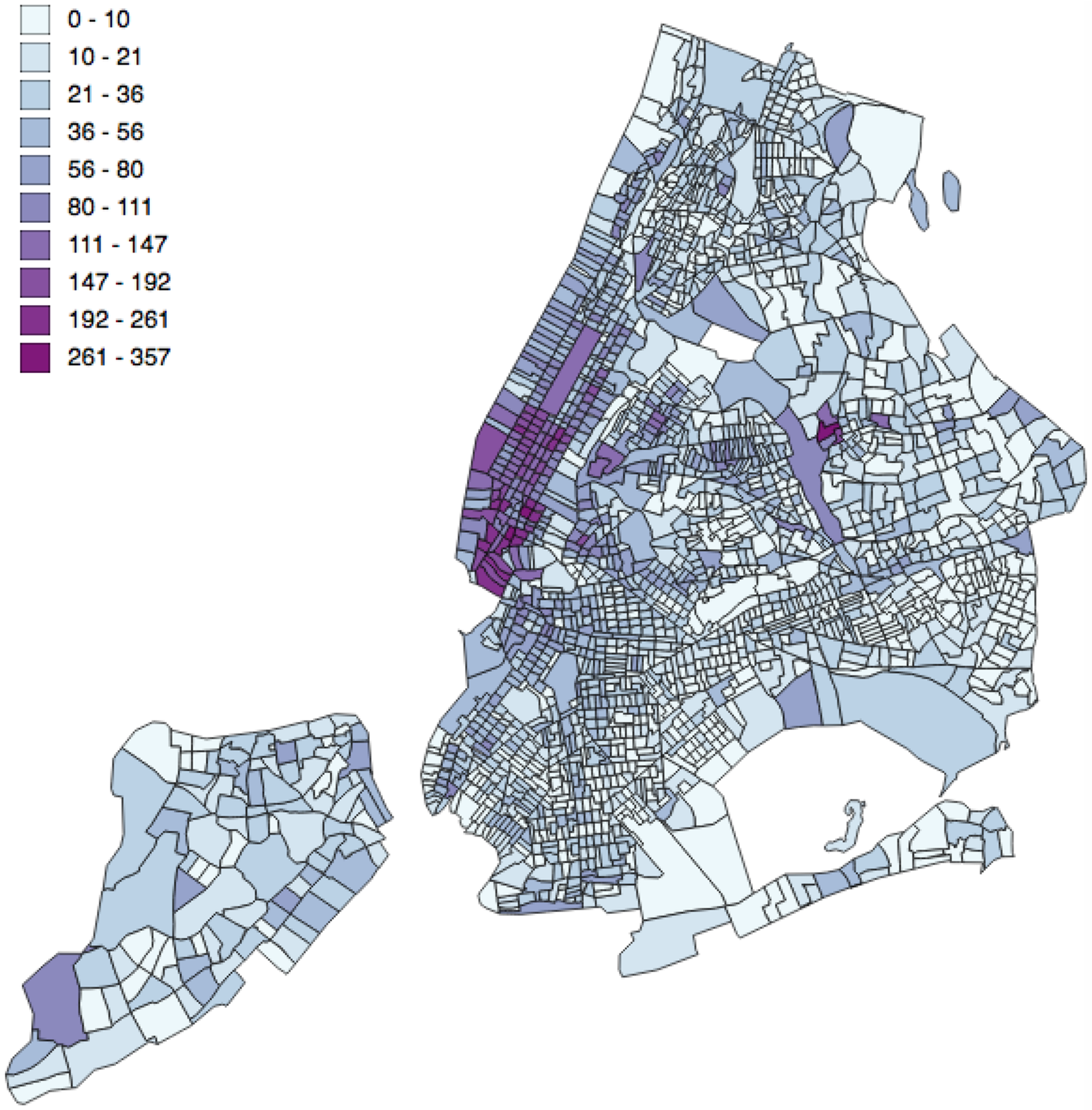, width=0.25\columnwidth}
\vskip 15pt
\psfig{file=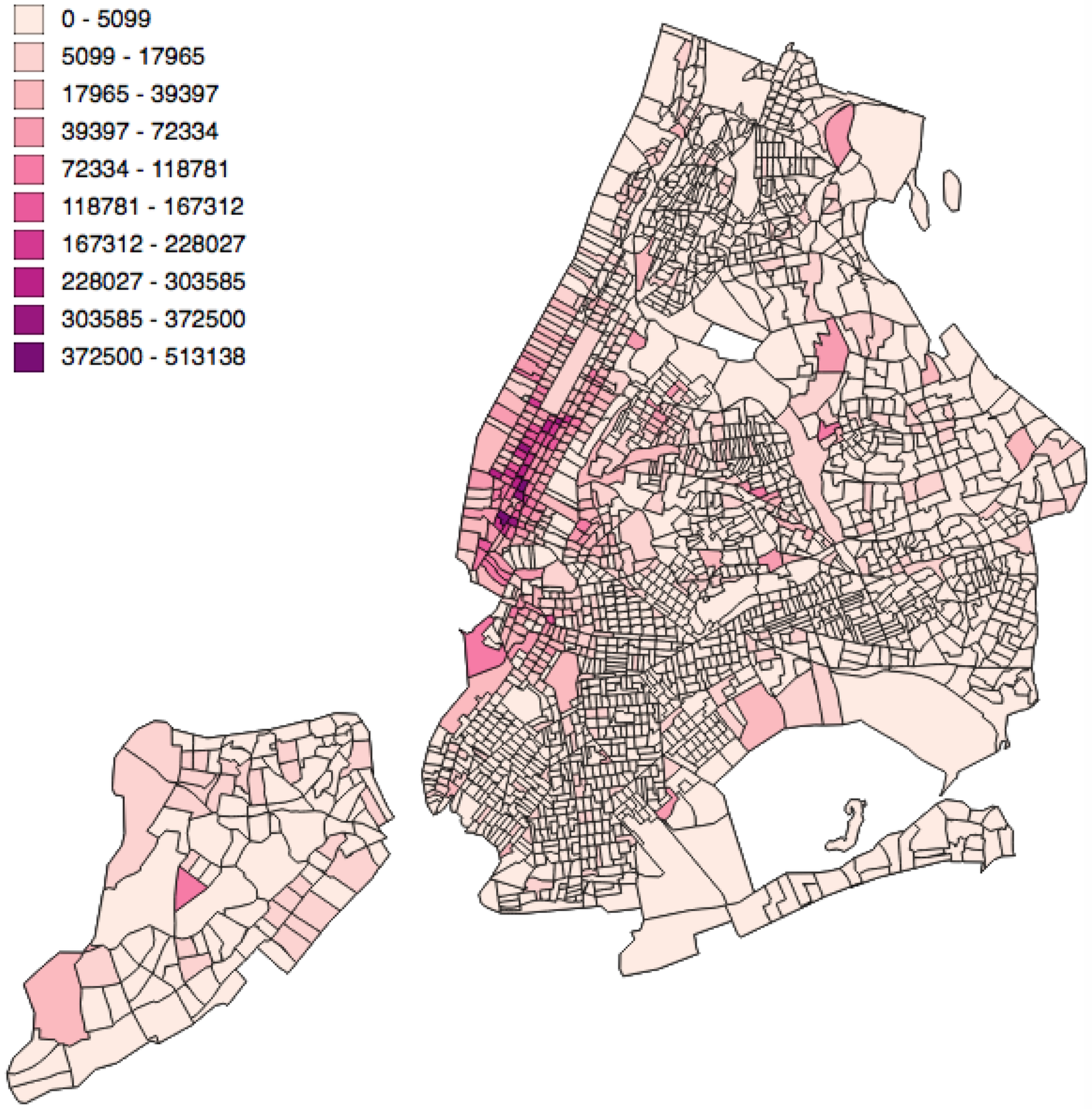, width=0.25\columnwidth}
\hskip 15pt
\psfig{file=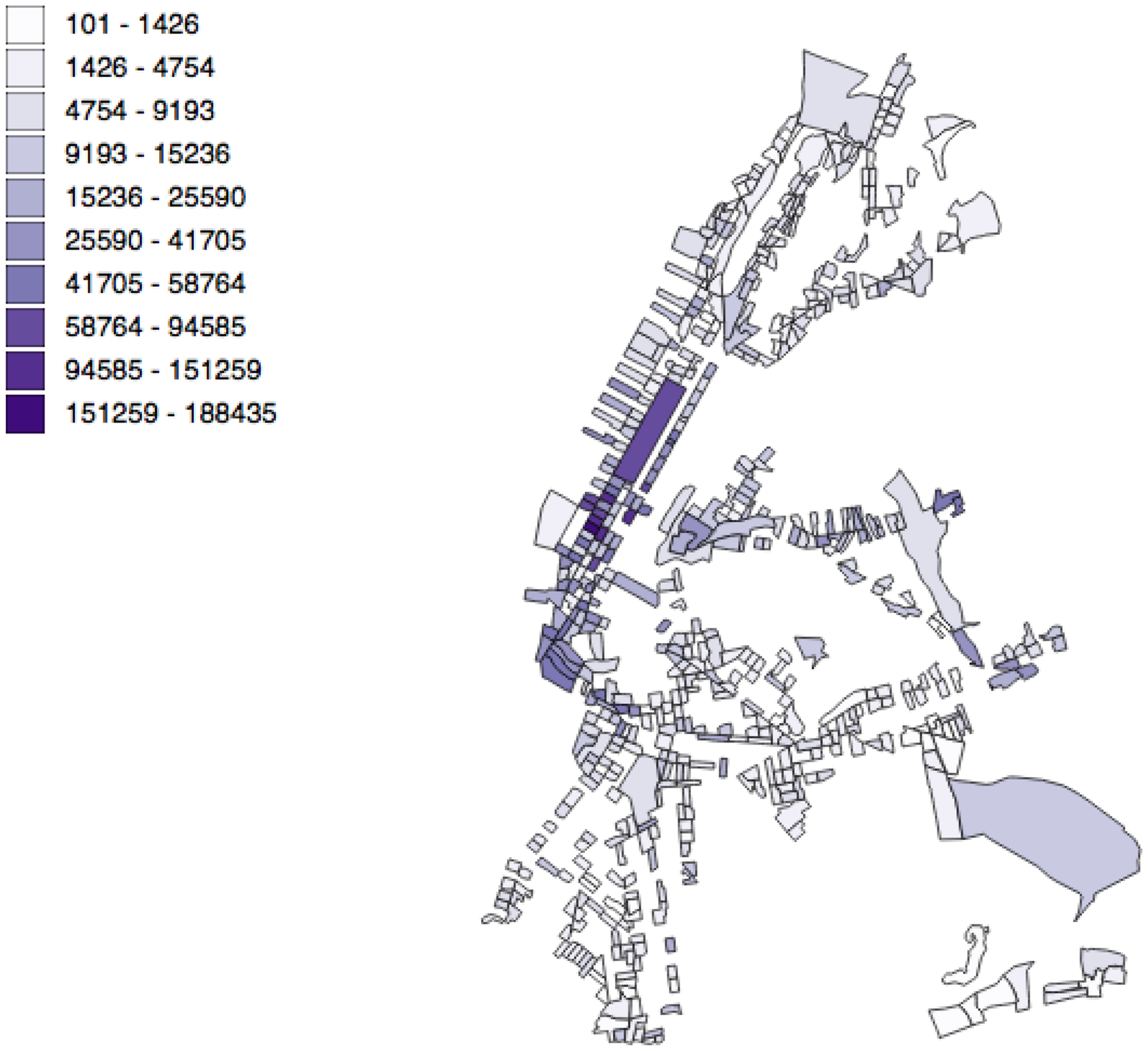, width=0.25\columnwidth}
\hskip 15pt
\psfig{file=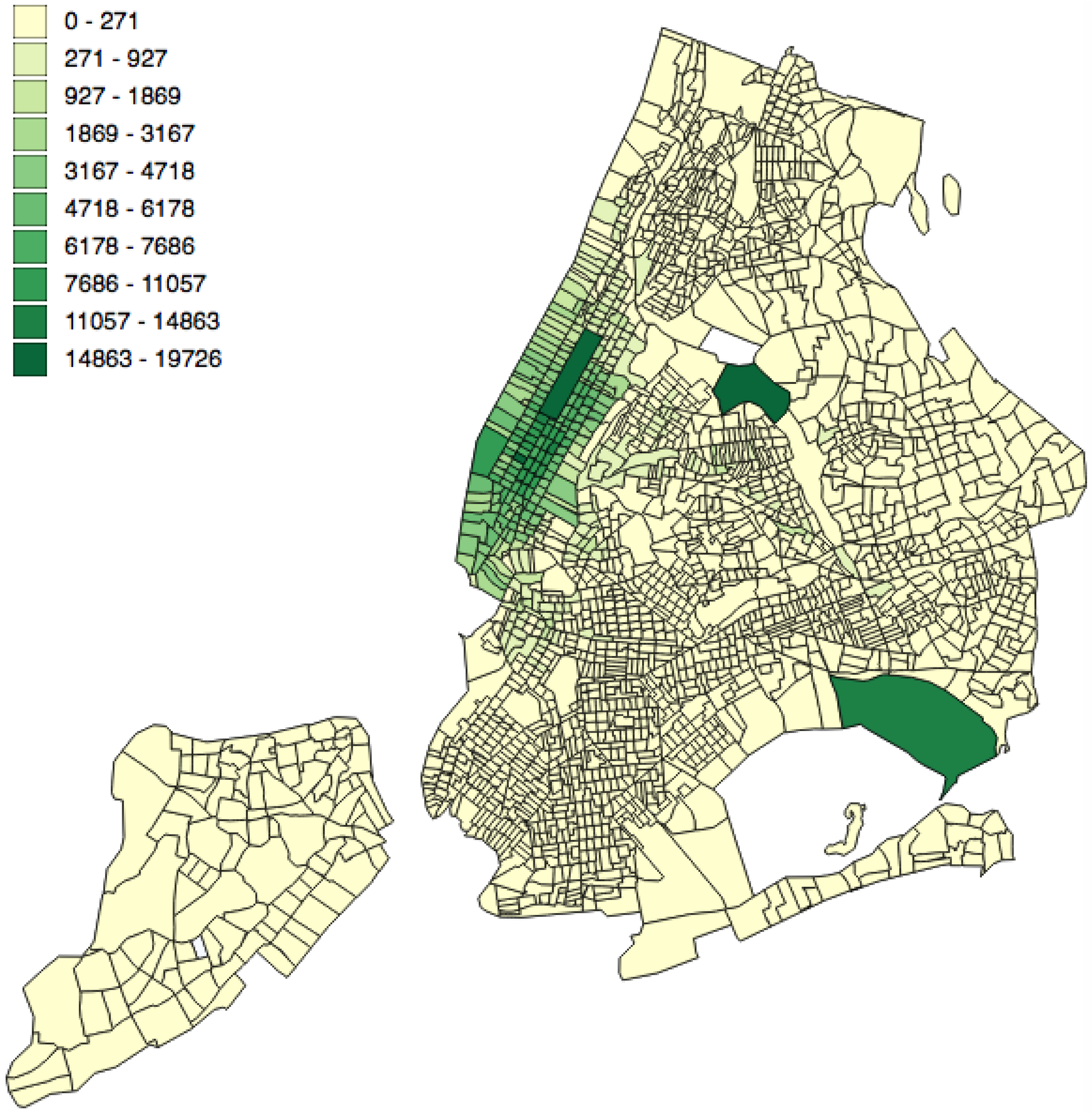, width=0.25\columnwidth}
\caption{NYC census tracts. From left to right and top to down: total number of 2015 incidents, percentage of rented houses, number of food venues, number of checkins in shops, number of 2015 subway exits (Mon-Fri average), number of 2015 picked up passengers (Mon-Fri average).} 
\label{fig:tracts}
\vskip -6pt
\end{figure}

Census tracts provide a stable set of geographic units for the presentation of statistical data and generally have a population size between 1,200 and 8,000 people, with an optimum size of 4,000 people\footnote{\url{https://www.census.gov/geo/reference/gtc/gtc_ct.html}}. In the case of NYC they span a few blocks and offer a natural unit for crime analysis at a detailed level. NYC has a total of 2,167 official census tracts. A few of these consist only of water or shoreline areas, which have not been experiencing any crime incidents in either of the analysis years. Furthermore, some NYC census tracts consist fully of military posts or jail facilities (like e.g. Fort Hamilton and Rikers Island) which exhibit different crime reporting schemes, next to restricted human presence. We remove these census tracts, and remain with a final of  $N = 2,154$ census tracts. Please note we still include many census tract with no resident population, like parks or airports, as these still experience crime, and now we have the possibility to model it by means of the ambient population measured by the alternative data sources. For visualization purposes, Figure~\ref{fig:tracts} depicts the 2015 aggregated crime counts per census tract, together with some example features computed at census tract level. All maps in this paper have been generated using the open source software QGIS\footnote{\url{http://www.qgis.org/en/site}}. 

\begin{figure} [h!]
\centering
\psfig{file=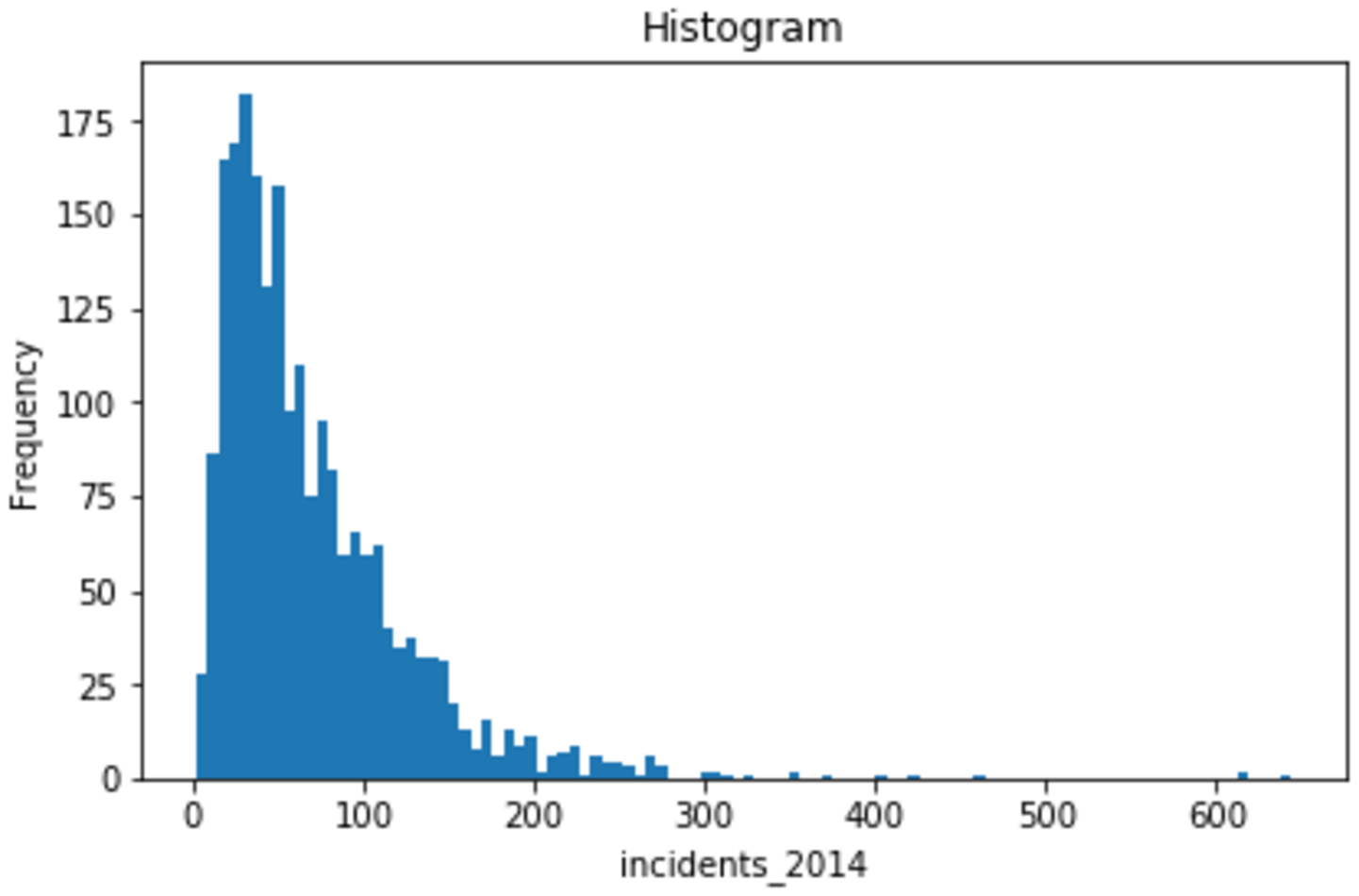, width=0.28\textwidth}
\hskip 75pt
\psfig{file=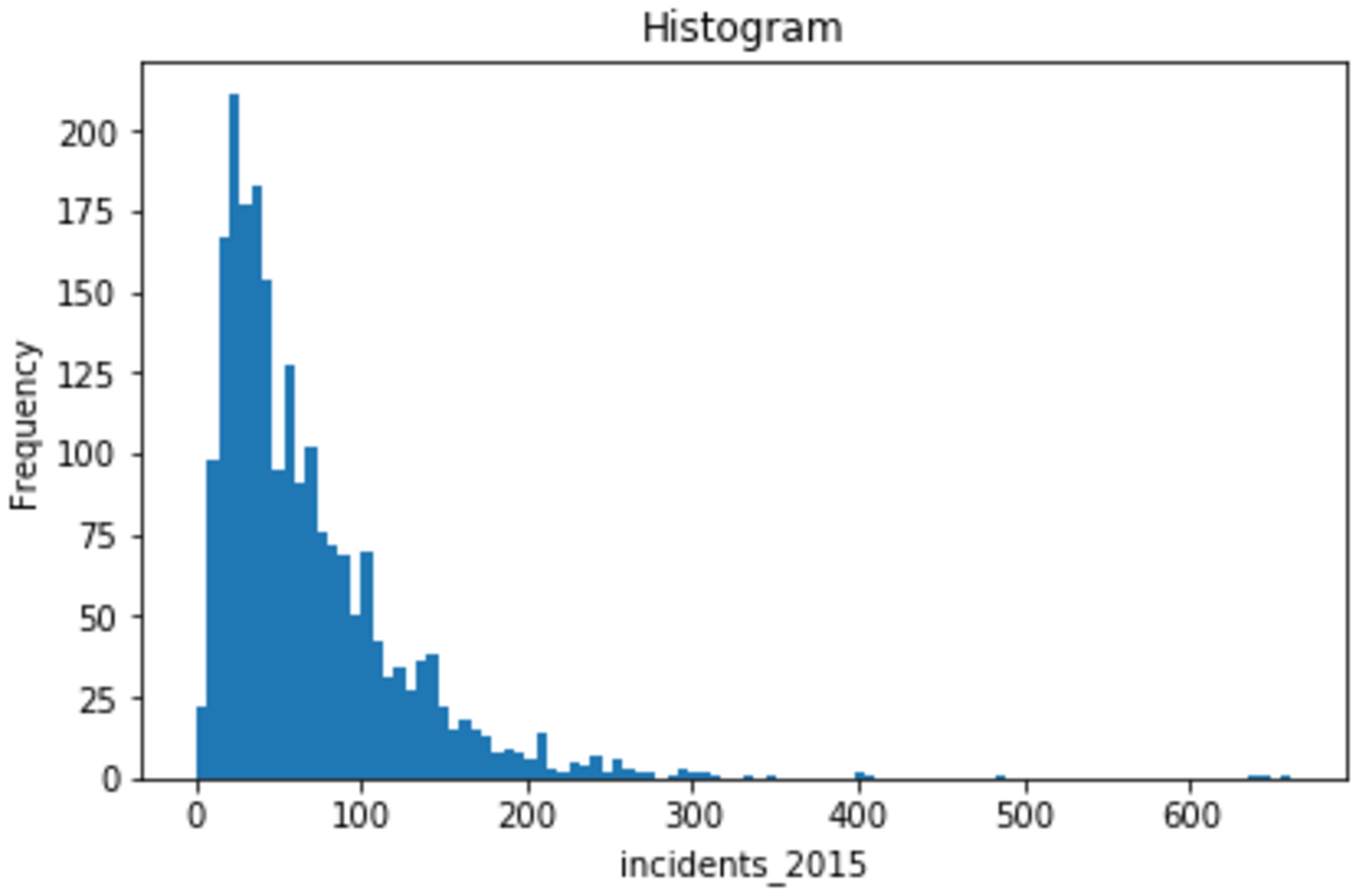, width=0.28\textwidth}
\psfig{file=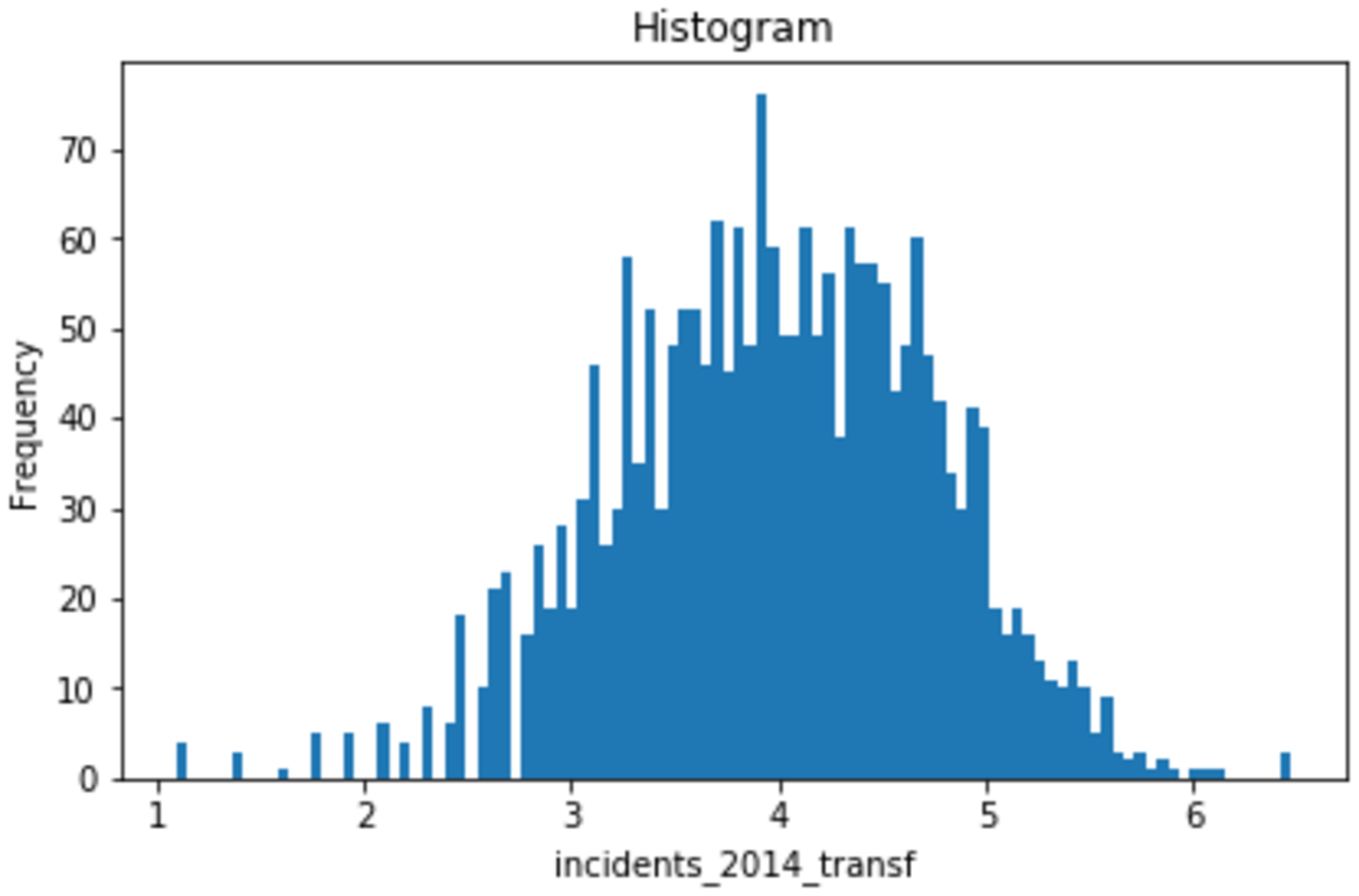, width=0.28\textwidth}
\hskip 75pt
\psfig{file=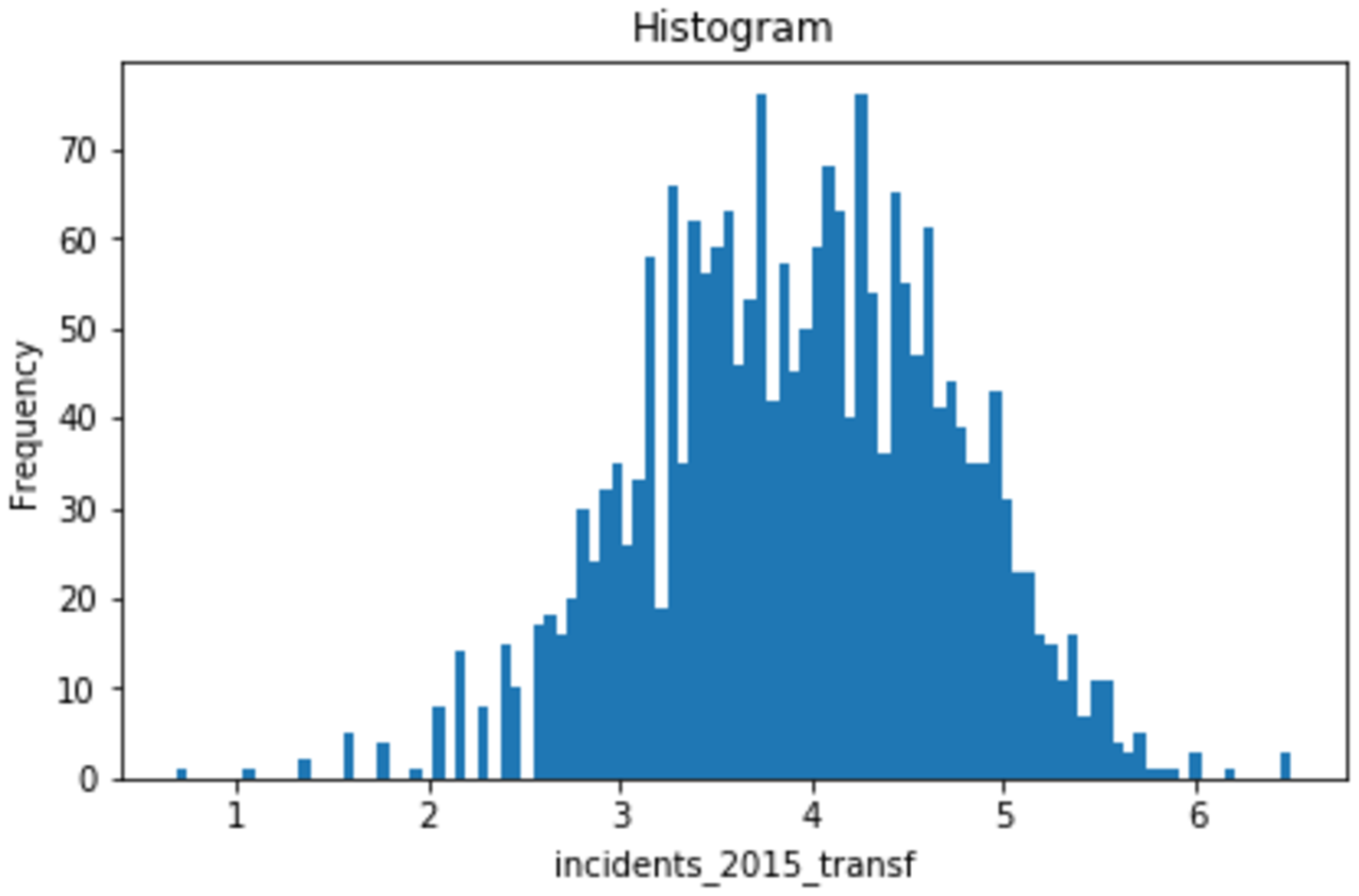, width=0.28\textwidth}
\caption{Histogram of the original and log-transformed total incidents per census track: 2014 (left) and 2015 (right).}
\label{fig:incidents_histogram}
\vskip -6pt
\end{figure}

Table~\ref{table:statistics} presents the descriptive statistics of crime counts of all types, while Figure \ref{fig:incidents_histogram} is depicting the histograms of the total incidents counts per census tract. We can observe that the distribution of the data is positively skewed with many observations having low count values. The various crime types expose also similar power law distributions, so for the prediction task below, we log-transform the dependent variable to correct for the positively skewed distribution, and use this as our dependent variable $y$.

\begin{table}[h!]
\scalebox{0.9}[0.9]{
\small
\begin{tabular}{ccccccc}
\hline
Incident type   &Min &Q1 &Median &Mean &Q3 &Max
\\ \hline
\textbf{\textit{2015}}\\
\hline
total incidents &1  &29 &52 &68  &91  
&661\\
grand larceny   &0  &10 &18 &28  &31  &519\\
robbery         &0  &4  &8  &12  &18  &90\\
burglary        &0  &4  &7  &9   &12  &90\\
assault         &0  &4  &8  &13  &19  &95\\
vehicle larceny &0  &2  &4  &5   &6  &38\\
\hline
\textbf{\textit{2014}}\\
\hline
total incidents &2  &31 &53 &70  &93  &644\\
grand larceny   &0  &11 &19 &29  &33  &512\\
robbery         &0  &3  &9  &12  &17  &67\\
burglary        &0  &5  &8  &10  &14  &83\\
assault         &0  &3  &8  &13  &19  &92\\
vehicle larceny &0  &2  &4  &5   &7   &61\\
\hline
\end{tabular}
}
\caption{Descriptive statistics of the crime data: counts per census tract for each year.}
\label{table:statistics}
\end{table}

\subsection{Prediction Features}
\label{sec:features}

In what concerns the independent variables $\boldsymbol{x}$, we craft an extensive set of features based on the collected massive datasets. Each feature represents a numeric score that characterizes a given census tract and is motivated by domain knowledge in criminology or urban computing, as explained below. We classify the features into three broad categories: (1) socio-demographic and economical features derived from the census sources, (2) spatial features which exploit solely the static information about the venues and subway stations, and (3) spatio-temporal features which integrate knowledge about the way the population moves around the city (by means of check-ins, subway entries/exits, taxi pick-ups/drop-offs). We have imported all data into a PostGIS-enabled Postgres database\footnote{\url{http://postgis.net/}}, which offers in-built optimized temporal and spatial queries that are required to process the data for feature generation per unit of analysis, as described in the remainder of this section. 



\subsubsection{Census Features}

To account for the fact that the units of analysis are heterogeneous, we include the census tract's \textbf{area} (in square miles) and \textbf{total population} as controls in the regression. 
We then proceed with a standard set of factors deemed in past criminological studies as significantly influential of crime and have been used also in related work in data mining, like \cite{Wang2016}. 

We start by operationalizing the concepts of the \textit{Lifestyle Exposure Theory} and \textit{Social Disorganization Theory}. We start with indicators of population at risk and of concentrated disadvantage \cite{Hindelang1978, Pratt2005, Sampson1997}: \textbf{fraction of male population}, \textbf{fraction of black population}, \textbf{fraction of hispanic population}, \textbf{fraction of population under the poverty level}. As violence has been associated with residential instability of neighborhoods \cite{Sampson1997, Graif2009}, we compute the \textbf{fraction of vacant households}, the \textbf{fraction of rented households} from the occupied ones, and the \textbf{fraction of stable population} (individuals who moved in prior to 2010). 

Furthermore, population diversity has been shown to play a role in the crime phenomenon \cite{Jacobs1961,Graif2009,Osgood2000} so we computed several diversity indexes based on the socio-demographic and economical information: a \textbf{racial ethnic diversity index}, an \textbf{age index}, and an \textbf{income diversity index}. The racial ethnic index is defined by the plurality of multiple ethnic and racial groups within a certain area and is computed based on five exhaustive and mutually exclusive aggregates (non-Hispanic whites, non-Hispanic blacks, Hispanics of any race, Asians, and others -- Native Americans, members of other races, and multi-racial persons) \cite{Lee2012}. The age index measures the variance in ages of the residents across four main age groups (under 18, 18-34, 35-64, and over 65 years), and the income index measures the variance in household income across three main income levels (low, medium, and high-income households) \cite{Lima2010}.

\subsubsection{Spatial Features}
This category of features describes the characteristics of a neighborhood, as captured by the spatial distribution of the Foursquare venues and subway stations within its perimeter. In general, the venues can be seen as \textit{crime attractors} -- particular places to which offenders are attracted because of the known opportunities for particular types of crimes \cite{Brantingham1995}.

The \textbf{number of venues of each category} measures the venues counts within a census tract and it is a static popularity metric of that area. 
The \textbf{fractions of venues of each category} capture the specifics of the life within a census tract, and it is an empirical metric for the functional decomposition of that particular area in the city. 
The \textbf{venues diversity index} is then a single measurement capturing the diversity of this decomposition. 
Inspired by \cite{Karamshuk2013}, we use the entropy measurement from information theory \cite{Shannon1948} as a diversity metric. Intuitively, the entropy quantifies the uncertainty in predicting the category of a venue that is taken at random from the area.The final formula models the normalized Shannon diversity index (also called the Shannon equitability index \cite{Sheldon1969}), which is the Shannon diversity index divided by the maximum diversity. For a given census tract $t_i$, we denote the count of included venues of category $c$ with $ V_c(t_i)$ and the total number of included venues with $ V(t_i)$ and formally define the venues diversity index of that census tract as follows (we employ smoothing by adding the constant 1 to the numerator and denominator to prevent zero divisions):
\[ - \sum_{c \in C} (\frac{ 1 + V_c(t_i)}{ 1+ V(t_i)} \times \ln{\frac{ 1 + V_c(t_i)}{ 1 + V(t_i)}}) / \ln{|C|} \]
The higher the index, the more heterogeneous the area is in terms of types of places, and following that, in terms of functions and activities of the neighborhood, whereas a least entropic area would indicate an area with a dominant function. For example, a census tract dominated by venues from the College and University category, would indicate a part of the city where people primarily study and would have a low diversity index. 

Motivated by the work in \cite{Venerandi2015}, we generate a metric called the \textbf{offering advantage} which denotes to what extent a particular neighborhood offers more venues of a particular category in comparison to the average neighborhood. Intuitively, the presence of one venue of an unpopular category, is more informative in profiling a neighborhood than the presence of one venue from a well-spread category. The offering advantage of category $c$ in each census tract $t_i$ of the total $N$ census tracts in NYC, is computed with the following formula:  
\[ \frac{ 1 + V_c(t_i)}{ 1 + V(t_i)} \times \frac{total\_venues}{\sum_{i=1}^{N} V_c(t_i)} \]
where $total\_venues$ is the number of total venues in NYC with an assigned category.

Finally, based on the MTA dataset, we compute the \textbf{total number of subway stations} within each census tract, to reflect whether the area is subject to high-volume population transit from other parts of the city.

\subsubsection{Spatio-temporal Features}
In this section, we derive metrics of human activity in that area. We compute, analog to the census data, metrics of density and diversity -- but, while the census features exploit information about the reported residential population, the human dynamics features are computed based on the ambient population, as measured by their usage of public venues and transportation. Overall, the features in this categories describe possible \textit{crime generators}. Crime generators produce crime by creating particular times and places that provide appropriate concentrations of people and other targets \cite{Brantingham1995}. These features can also be connected to the \textit{Routine Activity Theory}, as they model the activity nodes where motivated offenders meet vulnerable targets.

The \textbf{number of checkins per category} measure the popularity of the area. The empirically observed Foursquare checkins can be regarded as a more accurate measure of human activity than the traditional population density statistics from the census.

We further exploit Foursquare usage in each census tract, by looking at the popular hours of the venues (those times of the week where the venues experience most activity -- checkins, reviews, etc.) and compute the \textbf{number of venues that are popular in a typical morning, afternoon, evening or night -- split by weekdays and weekends} in each. These features give valuable information about the temporal break-down of human activity in the area.

We then compute, analog to the previous section, the \textbf{fraction of checkins of each category} in the area. These can be seen as measurements of the intensity of the different activity contexts in which the population engages. For instance, an area with many checkins in the Residence category would correspond to a residential neighborhood, which is very different to an entertainment district, that would in turn be characterized by a high number of checkins in the Food, Nightlife Spot, and Shop and Service categories. We proceed by computing the \textbf{checkins diversity index}, as an index of the distribution of human activity within the census tract. It can be seen, that the venues and checkins diversity indexes are the best operationalization of Jacobs' and Newman's concept of mixed land use.

Inspired by recent work on digital neighborhoods \cite{Anselin2015}, we compute \textbf{local quotients} of (digital) social activity within an area. Let $C(t_i)$ denote the total number of checkins and $P(t_i)$ the total population count within a census tract. We then compute the concentrations of checkins relative to the number of businesses and to the reference census population:
\[ \frac{ 1 + C(t_i)}{total\_checkins} \times \frac{total\_venues}{1 + V(t_i)} \]
\[ \frac{ 1 + C(t_i)}{total\_checkins} \times \frac{total\_population}{1 + P(t_i)} \]
where $total\_checkins$ denotes the number of total checkins in NYC, and $total\_population$ the total census population of NYC.
Neighborhoods with local quotients $>> 1$ can be regarded as (digital) hot spots, while neighborhoods with local quotients $<< 1$ can be regarded as (digital) deserts.

It can be observed, that the offering advantage and the local quotient metrics are both refined measures of the \textit{relative intensity of human activity} in an area as opposed to the whole city (one being based on the static distribution of the venues, and the other on the more dynamic distribution of the checkins).

To make use of the temporal dimension of the turnstile subway data, we aggregate it to \textbf{weekly averages of the number of individuals entering and exiting the subway stations -- split into Mon-Fri and Sat-Sun intervals}. We also compute a \textbf{subway rides diversity index}, by considering these four different categories: subway entries/exists in week/weekend.

Finally, we exploit the taxi ride data and computed \textbf{weekly averages of the number of passengers being picked up or dropped off in the census tracts -- split into Mon-Fri and Sat-Sun intervals}. Complementary to the popular hours of the venues, and the subway features, these features should give an additional indication of the average in- and out-flows of the population traveling to and from the area. Finally, we compute a \textbf{taxi rides diversity index}, by considering the numbers of pick-up/drop-off rides within the neighborhood.

Across all three feature categories, we end up with a total of 89 features. For exemplification purposes, Figure~\ref{fig:tracts} depicts a selection of the 2015 features computed at census tract level. Spearman correlation tests and linear regressions have revealed significant correlations between many of the features and the different $y$ variables -- see supplementary material (section Descriptive Statistics). We decided to keep them all for the following step, where the chosen machine learning algorithms, due to their internal structure, will be able to deal with higher number of (potentially correlated) features and rank them according to their predictive power.

We ought to acknowledge that other approaches to generating features would have been possible, all the way to completely automatically generating higher-level features from the raw data using techniques such as deep learning. We chose the middle way where we exploit a high number of features but use domain knowledge to generate them. This approach is prevalent in the urban computing and data science literature, used for instance: to identify optimal retail store placement \cite{Karamshuk2013}, to quantify the relationship between urban form and socio-economic indexes \cite{Venerandi2018}, or to understand economic behavior in the city \cite{Zhang2016}.
 
\section{ Results}
\label{sec:results}

\subsection{Model Evaluation}
We train three different tree-based machine learning models: a Random Forest regressor \cite{Breiman2001}, an Extra-Tree (Extremely Randomized Tree) regressor \cite{Geurts2006}, and a Gradient-Boosting regressor \cite{Friedman2002} -- all known in the literature for their ability to yield competitive prediction quality in high-dimensional heterogeneous feature spaces. Due to their non-parametric nature, they make no assumption about the data and can work with many, collinear features, while also requiring little preparation of the data \cite{Hastie2009}. On the other hand, linear models assume that the explaining variables are non-collinear, which is not the case in our data-rich setup. Furthermore, a linear model has proved to yield poor performance on our datasets and is not reported.

Random forests are very popular in practice, as they are easy to use, robust, and yield good performance. An entire set of decision trees are grown at training time, and their mean prediction is output at testing time, thus lowering the variance of the individual learners. The Extra-Trees add a third level of randomization in comparison to the random forests, in that the split tests at each node of the decision trees are random, next to the chosen sub-sets of samples and features. In practice, they yield sometimes better performance thanks to the introduced smoothing effect, and also remove computational burdens linked to the determination of optimal cut-points in random forests. While these first two models are averaging models and build their constituent decision trees in parallel, Gradient-Boosting builds the model in a stage-wise fashion. It constructs additive regression models by sequentially fitting a simple base learner on the current pseudo-residuals. Boosted trees have been shown to be the best performing models across a variety of tasks, at least in the pre-deep-learning era \cite{Caruana2006}. 

In addition, all these tree-based ensemble methods can be exploited to infer the relative importance of the input variables (based on the order in which they appear in the constituent decision trees) and to rank them accordingly \cite{Hastie2009}.

Internally, the regressors always optimize the mean squared error ($MSE$) total number of log-transformed incidents $y$, and we report two metrics: $MSE$, as well as the coefficient of determination ($R^2$). The $MSE$ metric is given by $\frac{1}{n}\sum_{i=1}^{n}{(y_i-\hat{y_i})}^2$, with lower scores being preferred. The $R^2$ metric measures the percentage of variance in the dependent variable that the model at hand explains: $1- \frac{\sum_{i=1}^{n}{(y_i-\hat{y_i})}^2}{\sum_{i=1}^{n}{(y_i-\bar{y_i})}^2}$, where $y_i$ are the true values, $\hat{y_i}$ are the predicted values, and $\bar{y_i}$ is the mean of the sample. Best possible score is $1.00$ and it can be negative (because the model can be arbitrarily worse). A constant model that always predicts the expected value of $y$, disregarding the input features, would get a score of $0.00$. It primarily helps us to compare models between the different feature configurations, but it can also be used to compare the performance on the different incident types, as it is independent of the sample range.  

We look at the performance of the algorithms across different model specifications, utilizing different subsets of the features introduced previously. The first model is a weak baseline consisting only of the socio-demographic and economical factors derived from the census sources. The second model is a strong baseline consisting additionally of the numbers of Foursquare venues/POIs per category. This model specification is designed to reproduce the nodal features from \cite{Wang2016}. We ought to note that the venues dataset might be slightly different from a standard dataset of POIs inferred for example from OpenStreetMap or Google Maps, as the Foursquare venues set is biased towards establishments where people spend time, and map already better to the concept of crime attractors then standard POIs. Hence, we expect that venues counts would outperform standard POI counts as features in crime prediction models. The third model is making use of all human dynamics features inferred from the mobility data sources, while the forth model is a full specifications, exploiting the complete set of features.

Furthermore, in the supplementary material, we create three further model specifications, where each makes use, additionally to the standard census features, of the full feature set of a given data source: Foursquare, subway rides, and yellow/green taxi rides. This enables a direct comparison of the ubiquitous data sources in terms of their predictive power for the crime domain -- in case in practice a model selection decision should be required.

For each machine learning model, incident type, and features subset combination, we estimate the performance of the algorithms on new unseen data. To asses their \textit{geographical} out-of-sample generalization, we do the following \textit{model evaluation} experiment using nested cross-validation. In a nested cross-validation, two cross-validation loops are performed: one outer loop to measure the prediction performance of the estimator and one inner loop to choose the best hyper-parameters of the estimator. We implement this approach with 5 outer loops for \textbf{model assessment} (i.e. setting the size of the test set to 20\%), and 2 inner loops for \textbf{model selection} (i.e. setting the size of the training and validation sets to 40\%, respectively). Table \ref{table:test_regressors} presents the final average $MSE$ and $R^2$ scores and standard deviations of the models on the left-out test subsets. The resulting scores are therefore unbiased estimates of the prediction score on new geographical samples. 
We also provide a \textit{temporal} evaluation of the approaches, by training a model on the complete 2014 data (with 5-fold CV for hyper-parameter tuning, i.e. \textbf{model selection}) and testing it on the unseen 2015 data for \textbf{model assessment}.

Across all experiments, the hyper-parameters optimized in the validation phase of the Random Forest and Extra-Trees are the number of trees in the ensemble (values ranging from 50 to 400) and the maximal depth (values ranging from one third, to one half, to the full set of features). The first parameter controls the model complexity, while the second controls the level of pruning of the trees, in other words performing regularization to avoid overfitting. For Gradient Boosting, we perform a grid search over the number of trees (values ranging from 100 to 400), the maximal depth (values ranging from 1 to 4), and also the learning rate (values ranging from 0.01 to 0.2). The models were implemented in Python v2.7, with the help of the scikit-learn\footnote{\url{http://scikit-learn.org/stable/}} and pandas\footnote{\url{http://pandas.pydata.org/}} libraries. The supplementary material (section Model Assessment) presents validation and learning curves of the employed models. The validation curves show that we have properly chosen the parameter ranges for hyper-parameter tuning. Also, the learning curves show that, in our case, the models keep improving with more data, so we should use all available samples.

\begin{table}[t!]
\makebox[\textwidth]{
\scalebox{0.65}[0.65]{
\small
\begin{tabular}{ccccccccc}
\hline
&\multicolumn{2}{c}{Census}
&\multicolumn{2}{c}{Census + POI}
&\multicolumn{2}{c}{Human Dynamics}
&\multicolumn{2}{c}{Census + Human Dynamics} \\
\hline
&\multicolumn{1}{c}{MSE}
&\multicolumn{1}{c}{$R^2$}
&\multicolumn{1}{c}{MSE}
&\multicolumn{1}{c}{$R^2$}
&\multicolumn{1}{c}{MSE}
&\multicolumn{1}{c}{$R^2$}
&\multicolumn{1}{c}{MSE}
&\multicolumn{1}{c}{$R^2$}\\
\hline
\textbf{\textit{2015}}\\
\hline
\textbf{Total incidents}\\
\hline   
Random Forest      &{0.58}$\pm${0.11} &{0.33}$\pm${0.19} &{0.46}$\pm${0.05} &{0.58}$\pm${0.07} &{0.55}$\pm${0.07} &{0.38}$\pm${0.20} &{0.44}$\pm${0.03} &{0.62}$\pm${0.03}\\   
Extra-Tree         &{0.57}$\pm${0.10} &{0.35}$\pm${0.16} &{0.45}$\pm${0.04} &{0.60}$\pm${0.06} &{0.55}$\pm${0.03} &{0.40}$\pm${0.07}  &{0.43}$\pm${0.03} &{0.63}$\pm${0.03}\\  
Gradient Boosting  &{0.57}$\pm${0.10} &{0.35}$\pm${0.16} &{0.44}$\pm${0.04} &{0.61}$\pm${0.06} &{0.57}$\pm${0.06} &{0.36}$\pm${0.08}  &\textbf{{0.42}$\pm${0.03}} &\textbf{{0.65}$\pm${0.03}}\\ 
\hline
\textbf{Grand larcenies}\\
\hline          
Random Forest      &{0.72}$\pm${0.17} &{0.14}$\pm${0.18} &{0.53}$\pm${0.05} &{0.52}$\pm${0.08} &{0.53}$\pm${0.05} &{0.52}$\pm${0.10} &{0.50}$\pm${0.04} &{0.57}$\pm${0.06}\\   
Extra-Tree     	   &{0.70}$\pm${0.15} &{0.18}$\pm${0.12} &{0.52}$\pm${0.05} &{0.53}$\pm${0.08} &{0.53}$\pm${0.05} &{0.52}$\pm${0.08} &{0.50}$\pm${0.04} &{0.57}$\pm${0.06}\\  
Gradient Boosting  &{0.71}$\pm${0.15} &{0.16}$\pm${0.13} &{0.53}$\pm${0.05} &{0.52}$\pm${0.08} &{0.53}$\pm${0.05} &{0.52}$\pm${0.08} &\textbf{{0.49}$\pm${0.03}} &\textbf{{0.59}$\pm${0.07}}\\ 
\hline
\textbf{Robberies}\\
\hline          
Random Forest      &{0.70}$\pm${0.05} &{0.36}$\pm${0.11} &{0.65}$\pm${0.05} &{0.46}$\pm${0.10} &{0.77}$\pm${0.06} &{0.23}$\pm${0.13} &\textbf{{0.62}$\pm${0.04}} &\textbf{{0.50}$\pm${0.08}}\\   
Extra-Tree     	&{0.69}$\pm${0.06} &{0.38}$\pm${0.12} &{0.64}$\pm${0.04} &{0.47}$\pm${0.07} &{0.77}$\pm${0.04} &{0.23}$\pm${0.10} &{0.62}$\pm${0.04} &{0.49}$\pm${0.08}\\  
Gradient Boosting  &{0.68}$\pm${0.05} &{0.40}$\pm${0.11} &{0.63}$\pm${0.05} &{0.48}$\pm${0.09} &{0.77}$\pm${0.03} &{0.22}$\pm${0.09} &{0.62}$\pm${0.04} &{0.49}$\pm${0.08}\\  
\hline
\textbf{Burglaries}\\
\hline        
Random Forest      &{0.60}$\pm${0.04} &{0.19}$\pm${0.03} &{0.55}$\pm${0.03} &{0.31}$\pm${0.05} &{0.62}$\pm${0.04} &{0.13}$\pm${0.12} &{0.56}$\pm${0.03} &{0.30}$\pm${0.06}\\   
Extra-Tree     	&{0.59}$\pm${0.04} &{0.21}$\pm${0.06} &{0.56}$\pm${0.03} &{0.31}$\pm${0.04} &{0.61}$\pm${0.03} &{0.16}$\pm${0.08} &{0.55}$\pm${0.04} &{0.31}$\pm${0.05}\\  
Gradient Boosting  &{0.57}$\pm${0.03} &{0.27}$\pm${0.04} &\textbf{{0.55}$\pm${0.03}} &\textbf{{0.32}$\pm${0.04}} &{0.63}$\pm${0.02} &{0.11}$\pm${0.06} &{0.56}$\pm${0.03} &{0.29}$\pm${0.04}\\ 
\hline
\textbf{Assaults}\\
\hline       
Random Forest      &{0.68}$\pm${0.04} &{0.46}$\pm${0.09} &{0.61}$\pm${0.03} &{0.56}$\pm${0.05} &{0.78}$\pm${0.05} &{0.27}$\pm${0.14} &{0.61}$\pm${0.03} &{0.56}$\pm${0.07}\\   
Extra-Tree         &{0.67}$\pm${0.02} &{0.47}$\pm${0.07} &\textbf{{0.60}$\pm${0.04}} &\textbf{{0.58}$\pm${0.05}} &{0.79}$\pm${0.03} &{0.27}$\pm${0.10} &{0.60}$\pm${0.03} &{0.58}$\pm${0.06}\\  
Gradient Boosting  &{0.66}$\pm${0.04} &{0.48}$\pm${0.07} &{0.61}$\pm${0.04} &{0.57}$\pm${0.06} &{0.80}$\pm${0.05} &{0.26}$\pm${0.08} &{0.60}$\pm${0.03} &{0.57}$\pm${0.07}\\  
\hline
\textbf{Vehicle larcenies}\\
\hline        
Random Forest      &{0.62}$\pm${0.08} &{0.10}$\pm${0.12} &{0.61}$\pm${0.07} &{0.12}$\pm${0.10} &{0.63}$\pm${0.03} &{0.04}$\pm${0.06} &\textbf{{0.58}$\pm${0.03}} &\textbf{{0.19}$\pm${0.04}}\\   
Extra-Tree     	&{0.61}$\pm${0.05} &{0.13}$\pm${0.06} &{0.62}$\pm${0.06} &{0.10}$\pm${0.05} &{0.64}$\pm${0.03} &{0.00}$\pm${0.10}  &{0.61}$\pm${0.05} &{0.12}$\pm${0.03}\\  
Gradient Boosting  &{0.62}$\pm${0.08} &{0.09}$\pm${0.12} &{0.61}$\pm${0.07} &{0.11}$\pm${0.08}  &{0.62}$\pm${0.02} &{0.07}$\pm${0.06} &{0.59}$\pm${0.04} &{0.16}$\pm${0.04}\\ 
\hline
\textbf{\textit{2014}}\\
\hline
\textbf{Total incidents}\\
\hline 
Random Forest   &{0.58}$\pm${0.10} &{0.29}$\pm${0.18} &{0.45}$\pm${0.06} &{0.57}$\pm${0.09} &{0.56}$\pm${0.06} &{0.35}$\pm${0.10} &\textbf{{0.44}$\pm${0.05}} &\textbf{{0.59}$\pm${0.06}}\\   
Extra-Tree     	&{0.58}$\pm${0.10} &{0.30}$\pm${0.17} &{0.45}$\pm${0.05} &{0.58}$\pm${0.09} &{0.57}$\pm${0.06} &{0.32}$\pm${0.09} &{0.44}$\pm${0.05} &{0.59}$\pm${0.06}\\  
Gradient Boosting  &{0.58}$\pm${0.08} &{0.29}$\pm${0.14} &{0.45}$\pm${0.05} &{0.58}$\pm${0.06} &{0.56}$\pm${0.08} &{0.34}$\pm${0.14}  &{0.45}$\pm${0.06} &{0.59}$\pm${0.08}\\ 
\hline
\textbf{Grand larcenies}\\
\hline
Random Forest      &{0.70}$\pm${0.15} &{0.13}$\pm${0.17} &{0.52}$\pm${0.07} &{0.52}$\pm${0.09} &{0.53}$\pm${0.06} &{0.49}$\pm${0.08} &{0.50}$\pm${0.06} &{0.56}$\pm${0.07}\\   
Extra-Tree     	&{0.69}$\pm${0.14} &{0.17}$\pm${0.13} &{0.51}$\pm${0.07} &{0.53}$\pm${0.08} &{0.54}$\pm${0.06} &{0.49}$\pm${0.06} &{0.50}$\pm${0.07} &{0.56}$\pm${0.06}\\  
Gradient Boosting  &{0.72}$\pm${0.16} &{0.09}$\pm${0.17} &{0.52}$\pm${0.08} &{0.52}$\pm${0.08} &{0.53}$\pm${0.07} &{0.49}$\pm${0.05} &\textbf{{0.49}$\pm${0.06}} &\textbf{{0.57}$\pm${0.05}}\\ 
\hline
\textbf{Robberies}\\
\hline
Random Forest      &{0.70}$\pm${0.04} &{0.35}$\pm${0.11} &{0.65}$\pm${0.06} &{0.44}$\pm${0.12} &{0.80}$\pm${0.05} &{0.16}$\pm${0.16} &{0.64}$\pm${0.05} &{0.47}$\pm${0.10}\\   
Extra-Tree         &{0.70}$\pm${0.04} &{0.36}$\pm${0.11} &{0.64}$\pm${0.06} &{0.47}$\pm${0.10} &{0.81}$\pm${0.05} &{0.13}$\pm${0.18} &{0.63}$\pm${0.05} &{0.48}$\pm${0.10}\\  
Gradient Boosting  &{0.69}$\pm${0.05} &{0.37}$\pm${0.12} &{0.64}$\pm${0.06} &{0.46}$\pm${0.11} &{0.81}$\pm${0.08} &{0.12}$\pm${0.25} &\textbf{{0.62}$\pm${0.04}} &\textbf{{0.50}$\pm${0.08}}\\ 
\hline
\textbf{Burglaries}\\
\hline
Random Forest      &{0.64}$\pm${0.04} &{0.19}$\pm${0.02} &{0.59}$\pm${0.03} &{0.30}$\pm${0.05} &{0.63}$\pm${0.05} &{0.21}$\pm${0.08} &{0.58}$\pm${0.03} &{0.31}$\pm${0.05}\\   
Extra-Tree         &{0.63}$\pm${0.03} &{0.20}$\pm${0.03} &{0.58}$\pm${0.02} &{0.32}$\pm${0.03} &{0.64}$\pm${0.05} &{0.18}$\pm${0.07} &{0.58}$\pm${0.03} &{0.32}$\pm${0.05}\\  
Gradient Boosting  &{0.61}$\pm${0.03} &{0.27}$\pm${0.01} &\textbf{{0.58}$\pm${0.03}} &\textbf{{0.33}$\pm${0.02}} &{0.64}$\pm${0.06} &{0.18}$\pm${0.09} &{0.58}$\pm${0.04} &{0.32}$\pm${0.05}\\ 
\hline
\textbf{Assaults}\\
\hline
Random Forest      &{0.70}$\pm${0.04} &{0.43}$\pm${0.08} &{0.64}$\pm${0.05} &{0.53}$\pm${0.09} &{0.84}$\pm${0.07} &{0.18}$\pm${0.15} &{0.64}$\pm${0.04} &{0.53}$\pm${0.08}\\   
Extra-Tree     	   &{0.69}$\pm${0.04} &{0.45}$\pm${0.07} &\textbf{{0.62}$\pm${0.03}} &\textbf{{0.56}$\pm${0.06}} &{0.86}$\pm${0.05} &{0.14}$\pm${0.12} &{0.62}$\pm${0.04} &{0.56}$\pm${0.07}\\  
Gradient Boosting  &{0.68}$\pm${0.04} &{0.47}$\pm${0.08} &{0.62}$\pm${0.03} &{0.56}$\pm${0.06} &{0.84}$\pm${0.06} &{0.19}$\pm${0.10} &{0.66}$\pm${0.04} &{0.50}$\pm${0.07}\\  
\hline
\textbf{Vehicle larcenies}\\
\hline 
Random Forest      &{0.61}$\pm${0.04} &{0.11}$\pm${0.09} &{0.62}$\pm${0.04} &{0.10}$\pm${0.08} &{0.63}$\pm${0.02} &{0.05}$\pm${0.05} &\textbf{{0.59}$\pm${0.03}} &\textbf{{0.17}$\pm${0.05}}\\   
Extra-Tree         &{0.62}$\pm${0.03} &{0.08}$\pm${0.05} &{0.63}$\pm${0.04} &{0.07}$\pm${0.08} &{0.63}$\pm${0.02} &{0.05}$\pm${0.05} &{0.60}$\pm${0.02} &{0.14}$\pm${0.03}\\  
Gradient Boosting  &{0.62}$\pm${0.05} &{0.08}$\pm${0.09} &{0.62}$\pm${0.04} &{0.08}$\pm${0.08} &{0.64}$\pm${0.02} &{0.03}$\pm${0.04} &{0.59}$\pm${0.03} &{0.16}$\pm${0.04}\\ 
\hline  
\end{tabular}
}
}
\caption{Geographical out-of-sample results of the regressors using different subsets of the features: for each year, repeatedly trained on 80\% of the census tracts, and tested on 20\% of the census tracts.}
\label{table:test_regressors}
\vskip -10pt
\end{table}

\subsubsection{Geographical Evaluation} 
Looking at the 2015 geographical prediction in Table \ref{table:test_regressors}, we observe that the novel behavioral features derived from the different data sources improve significantly the census-only and census + POI baselines for all incident types, with the exception of burglaries and assaults, where the models already saturate at the hard baseline of census + POI. 
For the total number of incidents we achieve a competitive $R^2$ score of 65\%, followed by the grand larcenies, robberies, and assaults categories with scores from 50\% to 59\%, while for burglaries and especially for vehicle larcenies the scores are lower. This can be explained by the fact that the latter categories of crime are not driven by the population characteristics, but by the characteristics of the target: house and car, respectively. As we do not include attributes of the built environment and of the stollen goods in the models, it was expected that these specific two categories would generally perform worse in comparison to the other categories.

For the total number of incidents, the best model on the full data set achieves scores of 65\%, which is 30 percentage points better than the best model in the weak-baseline and 4 percentage points better than the hard-baseline. But the highest improvement that we observe in comparison to the census-only baseline is in the case of grand larcenies: roughly 41 and 7 percentage points, respectively. This crime category includes different kinds of thefts, including pickpocketing. It was therefore expected that data describing the popularity of an area would be most informative, yet the improvement is spectacular. The weak baseline performs best for the assaults category. This category groups offenses that involve inflicting injury upon others, and it is already well explained by the collected socio-demographic and economical attributes of the neighborhood. 

Furthermore, for the case of total incidents and grand larcenies, we observe that models based solely on attributes of the ambient population outperform the models based on the classical demographic features -- and, in the case of grand larcenies, even reach performance levels comparable with those of the census + POI baseline. Finally, comparing the datasource-specific models (provided in supplementary material -- section Additional Model Specifications), we conclude that the census + FS consistently outperforms the census + subway and the census + taxi models -- with the exception of the vehicle larcenies crime category, which performs poorly across the board. Comparing the additional predictive power of the subway vs taxi rides, we notice a significant advantage of the taxi usage data in case of the grand larcenies category.

Inspecting the results for the 2014 geographical prediction, we deduce very similar insights: the full models for the total incidents, grand larcenies and the robberies categories perform best, with their absolute achieved MSE/$R^2$ scores being slightly bigger/lower than on the 2015 data.

\begin{table}[h!]
\makebox[\textwidth]{
\scalebox{0.65}[0.65]{
\small
\begin{tabular}{ccccccccc}
\hline
&\multicolumn{2}{c}{Census}
&\multicolumn{2}{c}{Census + POI}
&\multicolumn{2}{c}{Human Dynamics}
&\multicolumn{2}{c}{Census + Human Dynamics} \\
\hline
&\multicolumn{1}{c}{MSE}
&\multicolumn{1}{c}{$R^2$}
&\multicolumn{1}{c}{MSE}
&\multicolumn{1}{c}{$R^2$}
&\multicolumn{1}{c}{MSE}
&\multicolumn{1}{c}{$R^2$}
&\multicolumn{1}{c}{MSE}
&\multicolumn{1}{c}{$R^2$}\\
\hline
\textbf{Total incidents}\\
\hline        
Random Forest      &0.11 &0.82 &0.07 &0.88 
&0.09 &0.84 &0.07 &0.88\\   
Extra-Tree         &0.11 &0.82 &0.07 &0.89 
&0.08 &0.87 &\textbf{0.07} &\textbf{0.89}\\ 
Gradient Boosting  &0.22 &0.64 &0.09 &0.85 
&0.12 &0.80 &0.08 &0.87\\ 
\hline
\textbf{Grand larcenies}\\
\hline          
Random Forest      &0.19 &0.73 &0.14 &0.81 
&0.14 &0.81 &\textbf{0.13} &\textbf{0.82}\\ 
Extra-Tree         &0.21 &0.71 &0.14 &0.81 
&0.14 &0.80 &0.14 &0.80\\ 
Gradient Boosting  &0.28 &0.61 &0.17 &0.77 
&0.16 &0.78 &0.15 &0.79\\ 
\hline
\textbf{Robberies}\\
\hline          
Random Forest      &0.27 &0.71 &0.24 &0.75 
&0.28 &0.70 &\textbf{0.23} &\textbf{0.75}\\ 
Extra-Tree         &0.26 &0.72 &0.23 &0.75 
&0.27 &0.70 &0.27 &0.71\\ 
Gradient Boosting  &0.38 &0.59 &0.29 &0.69 
&0.32 &0.66 &0.28 &0.70\\   
\hline
\textbf{Burglaries}\\
\hline        
Random Forest      &0.25 &0.47 &0.24 &0.50 
&0.25 &0.47 &0.24 &0.49\\   
Extra-Tree         &0.25 &0.47 &0.24 &0.50 
&0.33 &0.32 &0.32 &0.34\\ 
Gradient Boosting  &0.30 &0.38 &0.25 &0.47 
&0.27 &0.42 &\textbf{0.23} &\textbf{0.51}\\ 
\hline
\textbf{Assaults}\\
\hline       
Random Forest      &0.24 &0.75 &0.22 &0.77 
&0.28 &0.72 &0.22 &0.77\\ 
Extra-Tree         &0.24 &0.76 &\textbf{0.22} &\textbf{0.78} 
&0.28 &0.71 &0.27 &0.73\\ 
Gradient Boosting  &0.34 &0.65 &0.29 &0.71 
&0.46 &0.53 &0.24 &0.76\\ 
\hline
\textbf{Vehicle larcenies}\\
\hline        
Random Forest      &0.31 &0.31 &0.29 &0.34 
&0.31 &0.31 &0.30 &0.34\\ 
Extra-Tree         &0.33 &0.27 &\textbf{0.29} &\textbf{0.36} 
&0.37 &0.16 &0.38 &0.15\\
Gradient Boosting  &0.32 &0.28 &0.31 &0.30 
&0.34 &0.23 &0.30 &0.33\\ 
\hline
\end{tabular}
}
}
\caption{Temporal out-of-sample results of the regressors using different subsets of the features: trained on 2014 and tested on 2015.}
\label{table:pred_regressors}
\vskip -10pt
\end{table}

\subsubsection{Temporal Evaluation} 
Switching to the temporal prediction presented in Table \ref{table:pred_regressors} and in supplementary material (section Additional Model Specifications), we can observe that predicting future crime aggregates within the same neighborhoods appears to be easier than predicting crime aggregates in new neighborhoods as the ecological attributes of a neighborhood, as well as the aggregated crime levels, do not vary that much between the two years. The total number of incidents proves to be the most predictable from one year to the other -- with an $R^2$ score of 89\%. In terms of crime sub-types: grand larcenies, robberies and assaults remain the types that can be best predicted by the data. Similarly to the geographical evaluation, the human dynamics only models outperform the census only models in the case of total incidents and grand larcenies. With the exception of the census baseline, all model specifications including ubiquitous data perform similarly good, whereby the models including FS-derived features (census + POI, census + FS, and the full model) achieve the highest absolute scores. 

\subsection{Model Interpretation}

We now turn to \textbf{model interpretation}, where the focus will be (1) on examining the importance and the contribution of the individual features defined in Section \ref{sec:features} and (2) on understanding where in the city do the ambient population features improve the baseline models. 

\subsubsection{Feature Importance} 
This exercise will return those features that proved to be most discriminative for geographical crime prediction task. By examining them, we will be able to understand what type of factors are most relevant for the predictive algorithms, and also identify those criminological theories that have informed the best features. It is important to stress the fact that, these techniques would not allow us to infer any causal relationships between the features and the crime counts. The identified factors are most discriminative in the context of the used model, but they not necessarily best explain crime levels. 

\begin{figure}[t!] 
\centering
\psfig{file=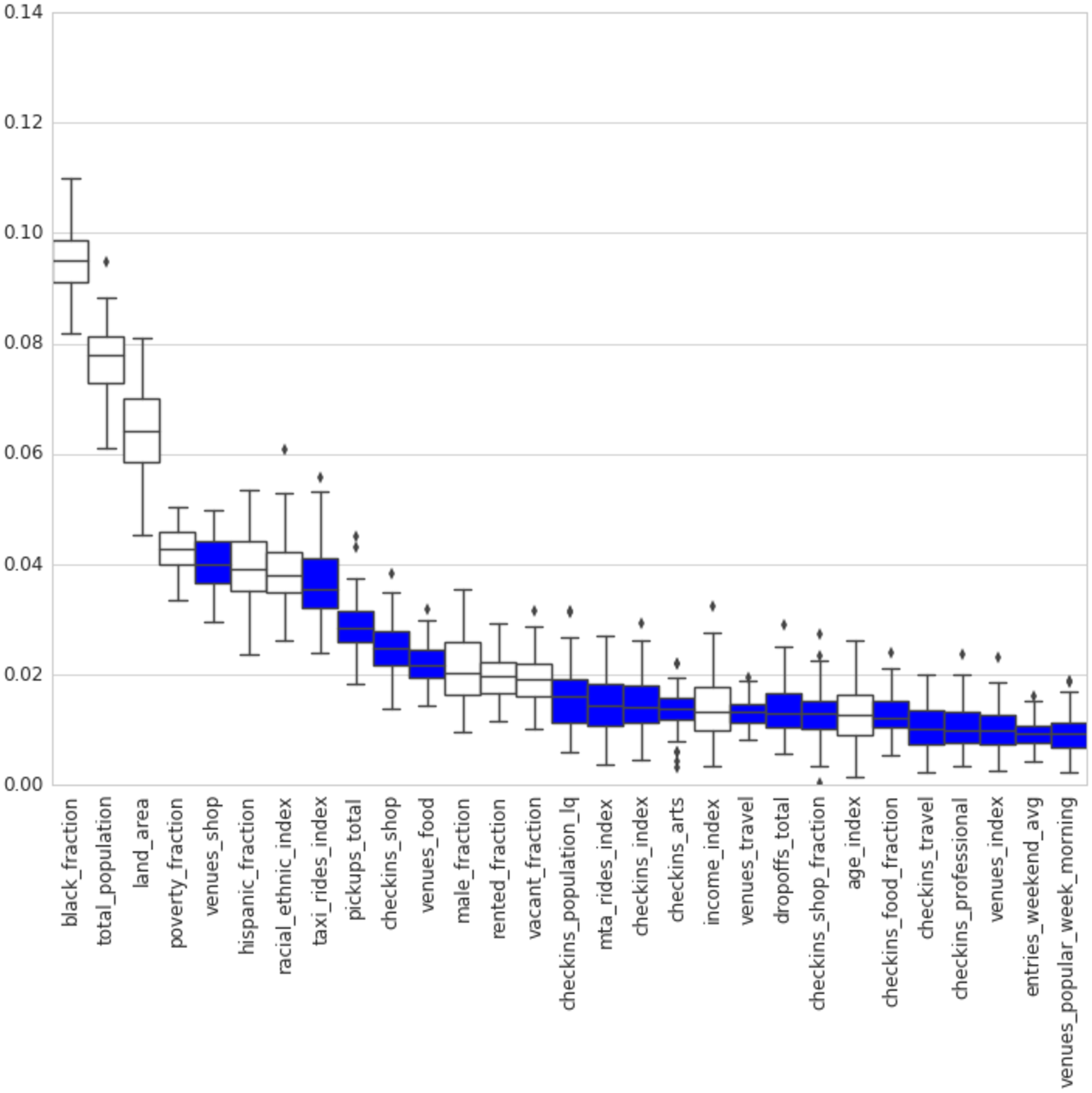, width=0.3\columnwidth}
\psfig{file=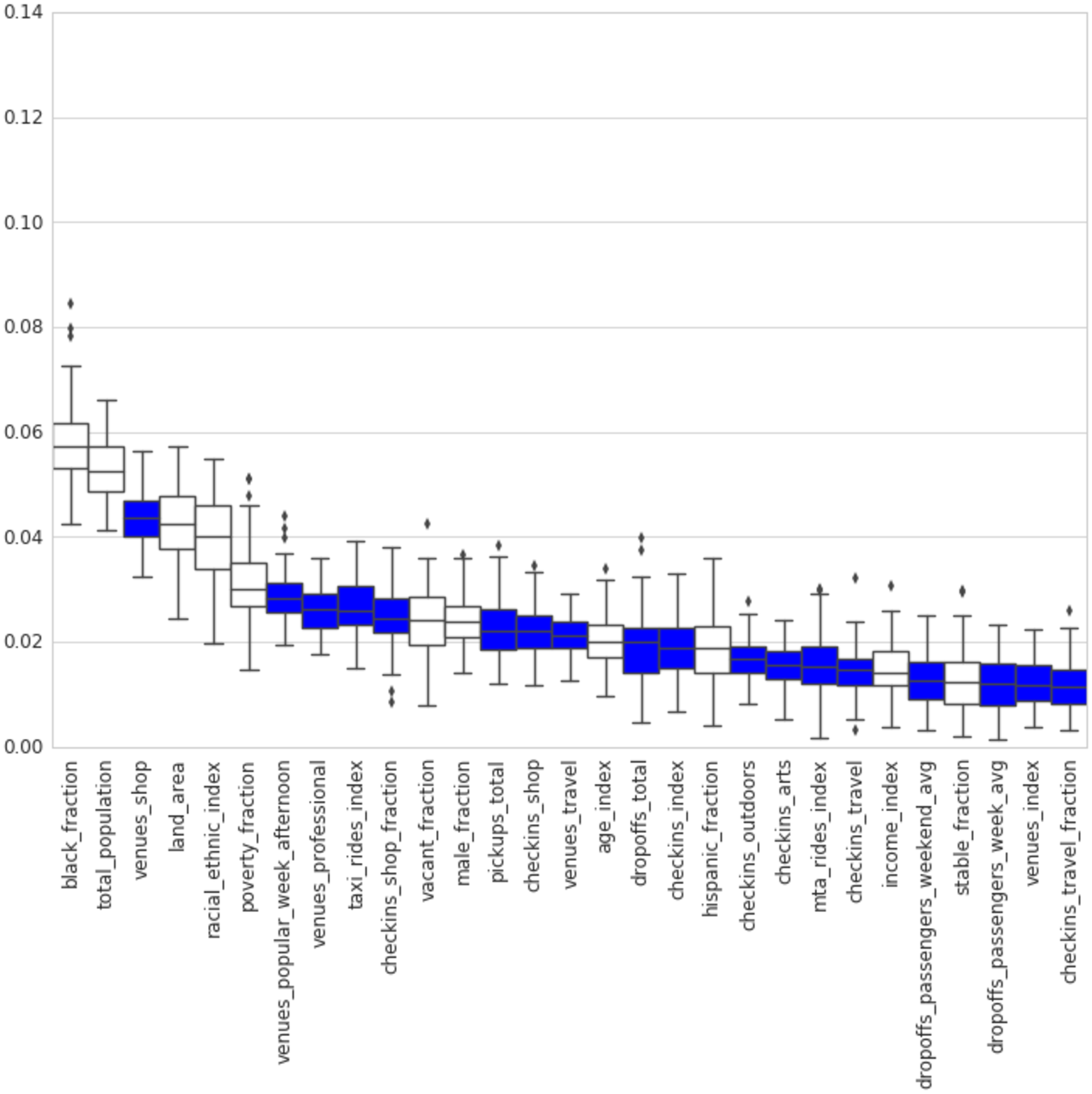, width=0.3\columnwidth}
\psfig{file=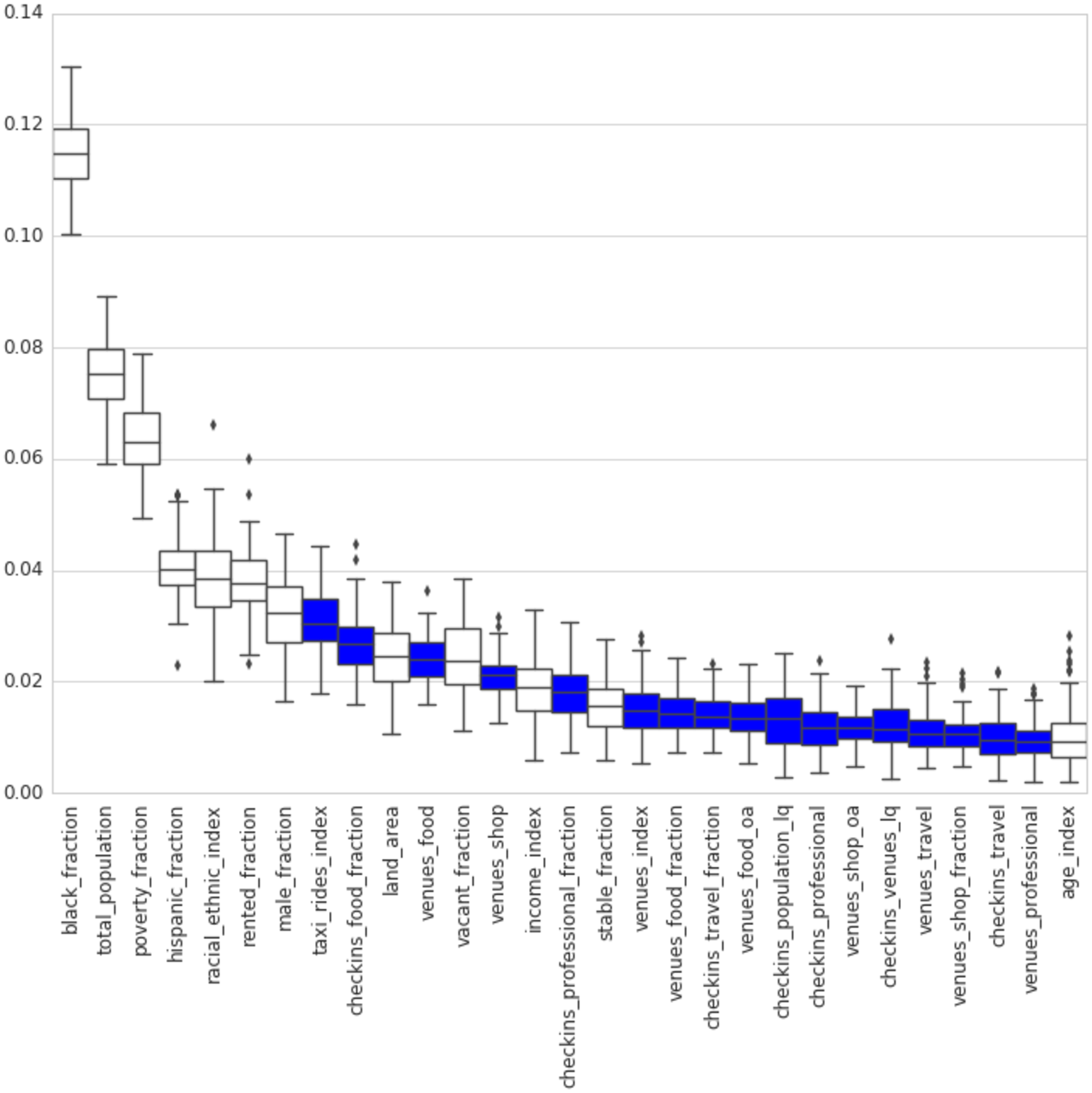, width=0.3\columnwidth}
\caption{Variable importance plots (top one third of the variables) reported by the Gradient Boosting Models (full specification). From left to right: 2015 total incidents, 2015 grand larcenies, and 2015 assaults.}
\label{fig:vip}
\vskip -6pt
\end{figure}

The supplementary material (section Feature Importances across Models) provides a complete view of the feature importances plots of all machine learning models, while here we concentrate on providing a stable ranking of the features within the most adequate model for this task: Gradient Boosting.
To test the stability of the features rank, we perform following bootstrapping procedure: we calculate the importance of the features for 100 random different samples (80\% of the data) and provide a box-plot ranked by the median importance of the outputs returned by the different samples. Figure \ref{fig:vip} visualizes the top one third variables in these rankings: in white features inferred from the census, in blue features inferred from human mobility data.

The traditional census features score indeed high across all three crime categories and across all algorithms. Specifically, we observe their very high contribution in the assaults model. As already hinted in the previous section, this type of violent crime remains best predicted by the attributes of the residential population in an area.

Also the spatial features from Foursquare have significant contributions across all models. The shopping venues contribute most in the grand larcenies category, followed by professional and travel venues. On the other hand, the food establishments, followed by the shopping establishments have a significant contribution in the assaults models. 

In terms of spatio-temporal features from Foursquare, we see importance assigned to many features derived from checkins data, like checkins in food and shops and checkins diversity index. We also see that the number of afternoon popular venues during the week receives a high weight for the grand larcenies category.

In terms of human dynamics features inferred from the taxi data, we notice especially high loadings for the diversity index of the taxi drives and the total number of pickups and for the  in the larcenies and total incidents categories.
The human dynamics features inferred from the subway data have in general a lower predictive contribution, with the diversity index ahaving the relative higher scores in this features subgroup and making it into the top features for total incidents and grand larcenies.

\subsubsection{Partial Dependence Plots} 

The above feature importance rankings only tell us \textit{which} features are predictive of crime, but not \textit{how} they contribute to the models. There are several approaches on how to achieve that. One approach is to plot partial dependency plots of the gradient boosting learners, another approach is to fit simple decision trees on the top discriminative features of the full models and extract prediction rules.

\begin{figure} [t!]
\centering
\psfig{file=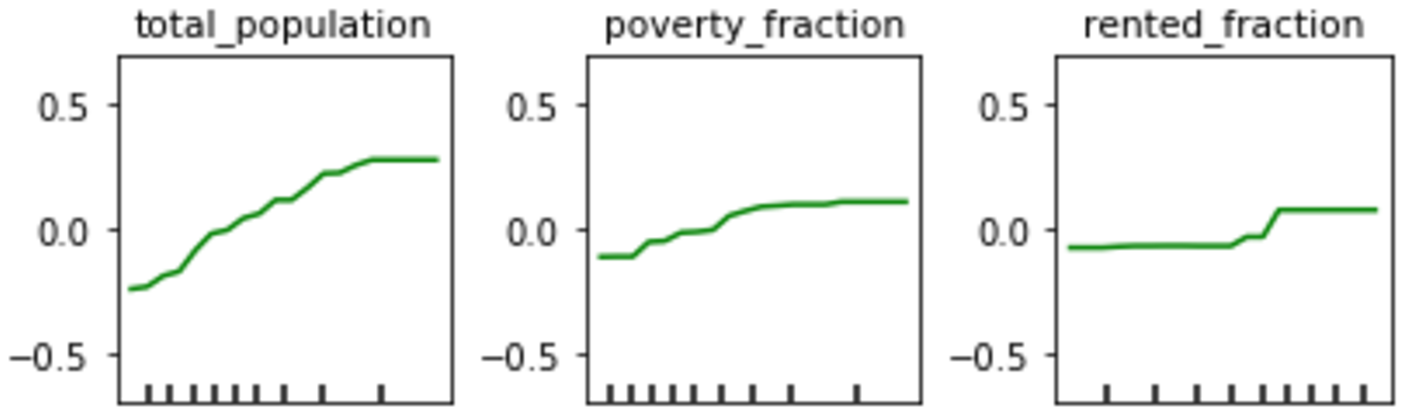, width=0.4\columnwidth}
\hskip 100pt
\psfig{file=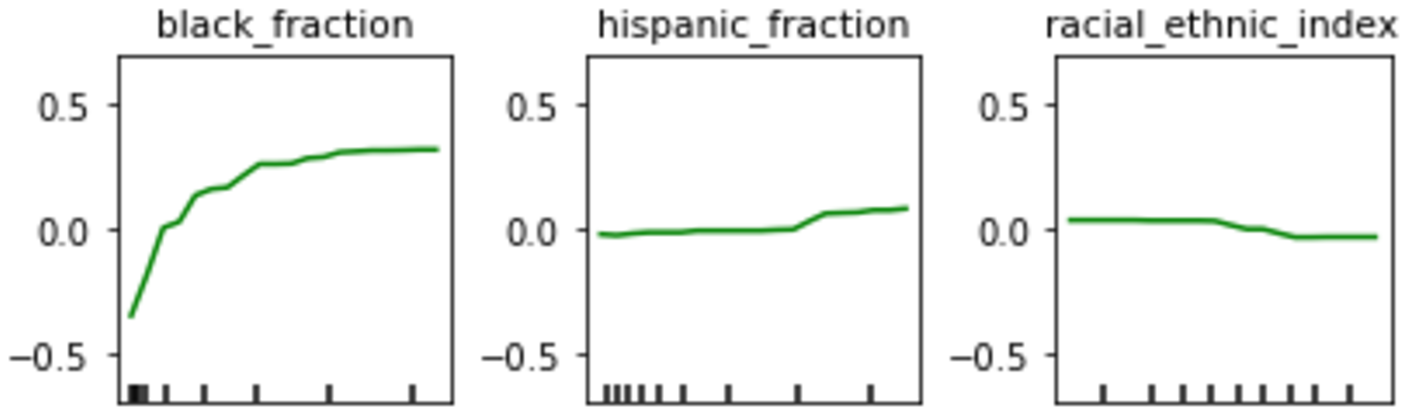, width=0.4\columnwidth}
\hskip 100pt
\psfig{file=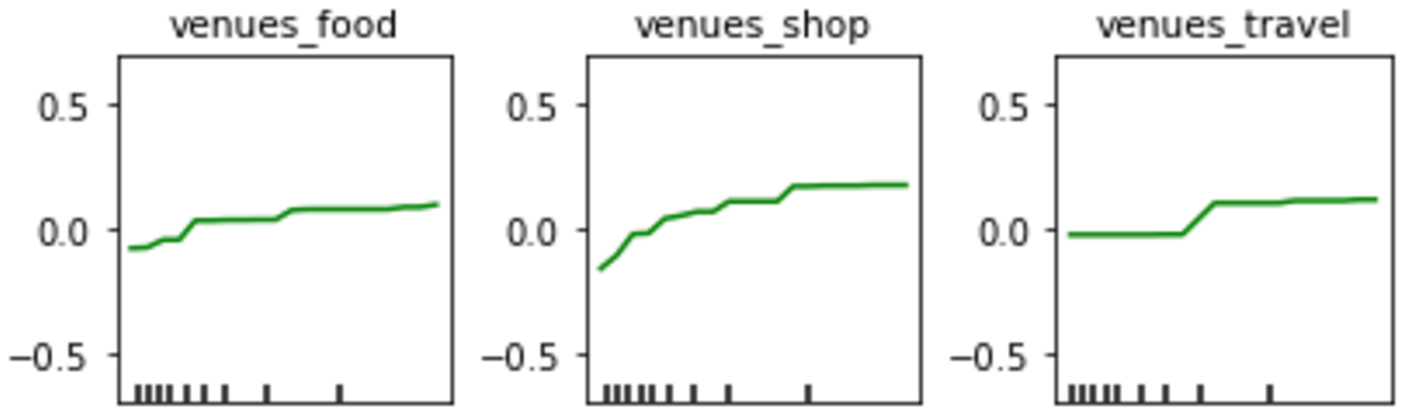, width=0.4\columnwidth}
\hskip 100pt
\psfig{file=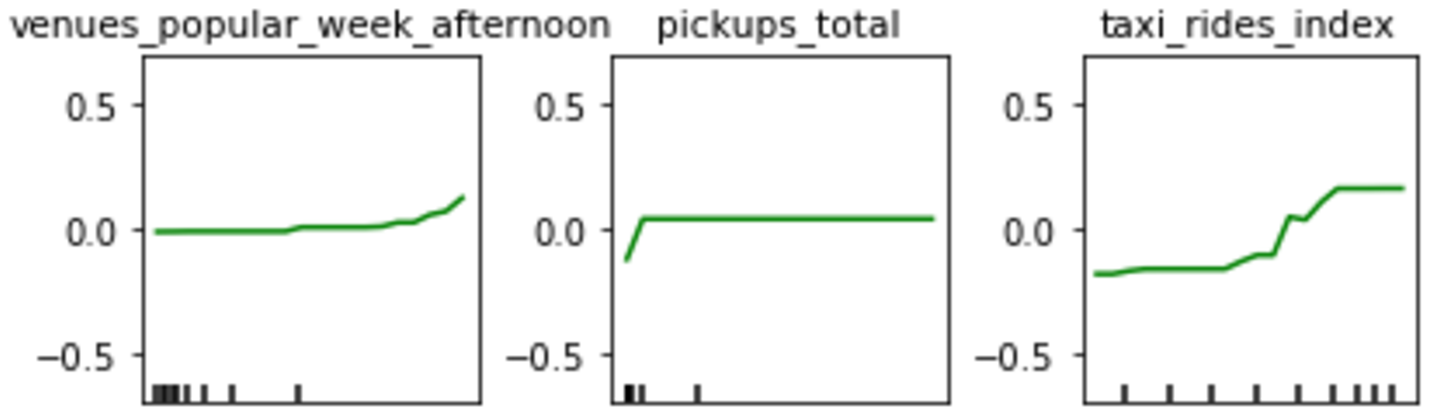, width=0.4\columnwidth}
\vskip 20pt

\psfig{file=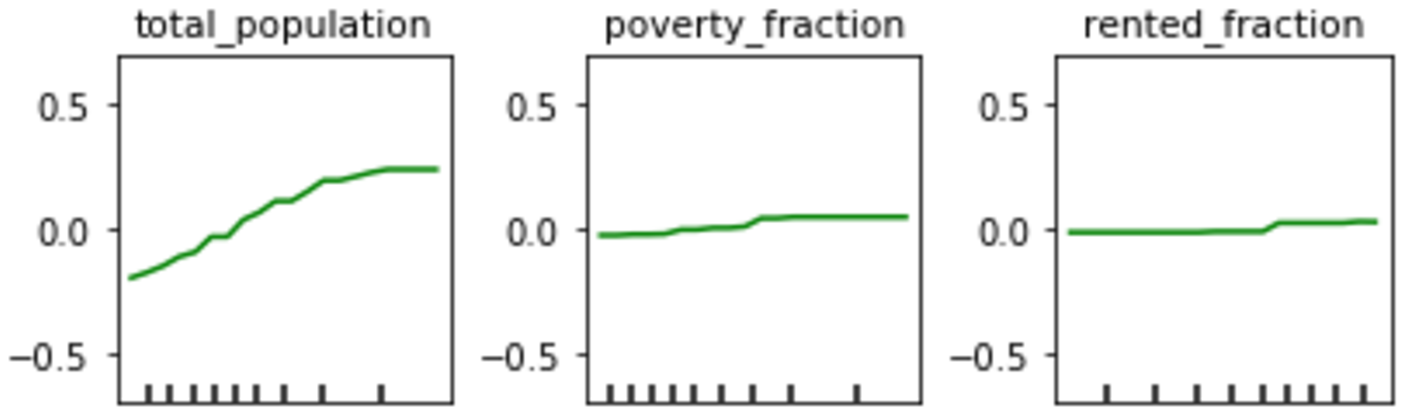, width=0.4\columnwidth}
\hskip 100pt
\psfig{file=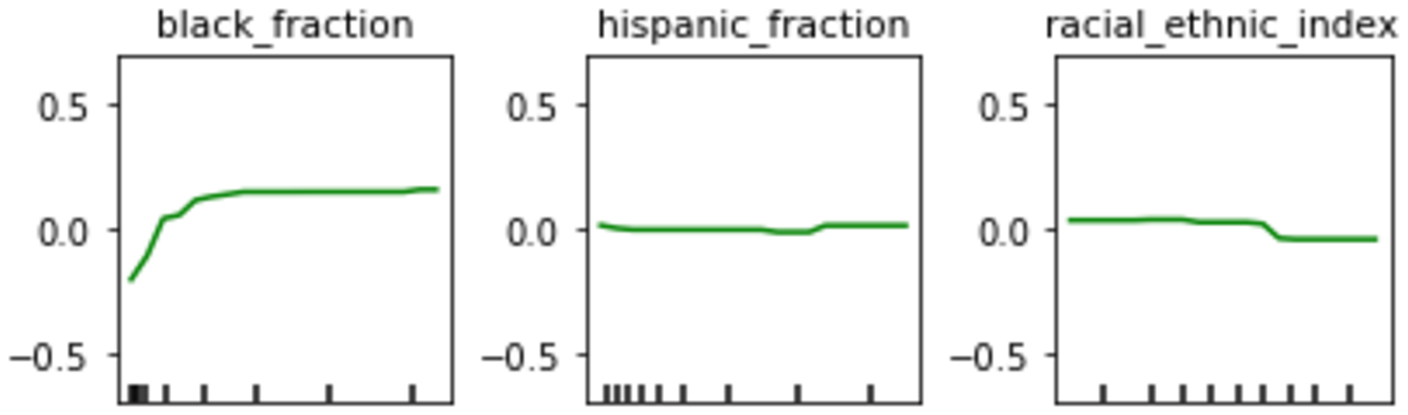, width=0.4\columnwidth}
\hskip 100pt
\psfig{file=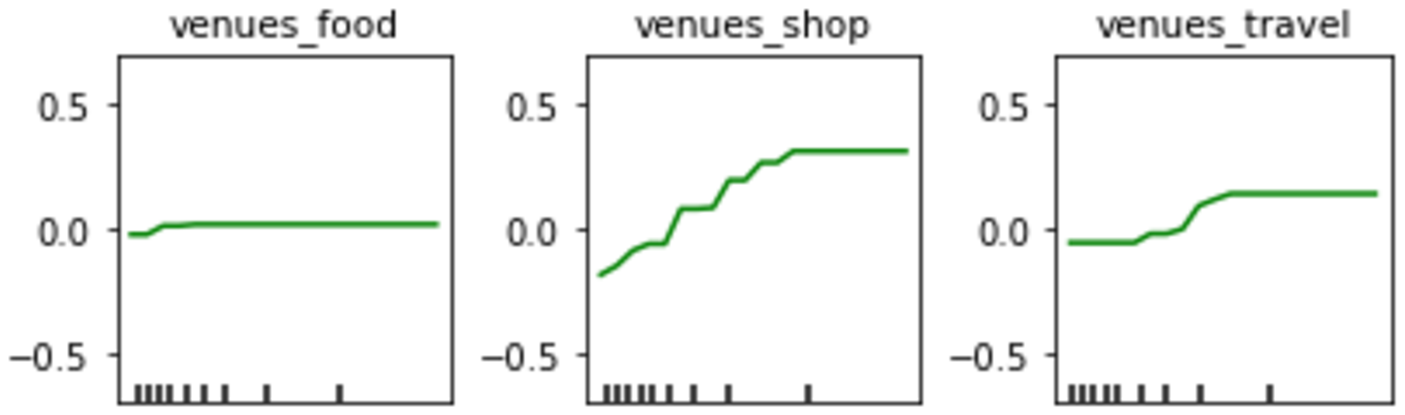, width=0.4\columnwidth}
\hskip 100pt
\psfig{file=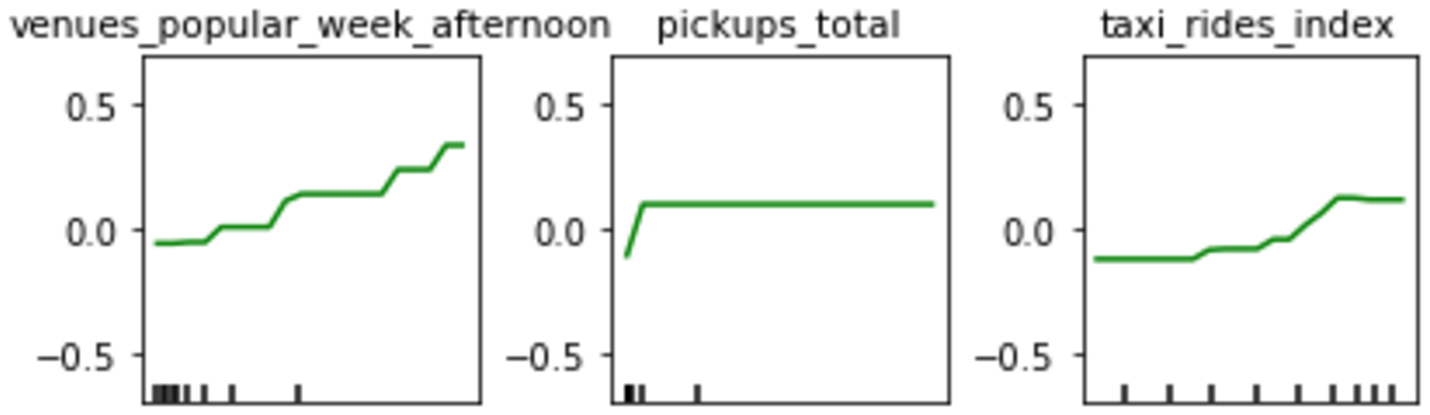, width=0.4\columnwidth}
\vskip 20pt

\psfig{file=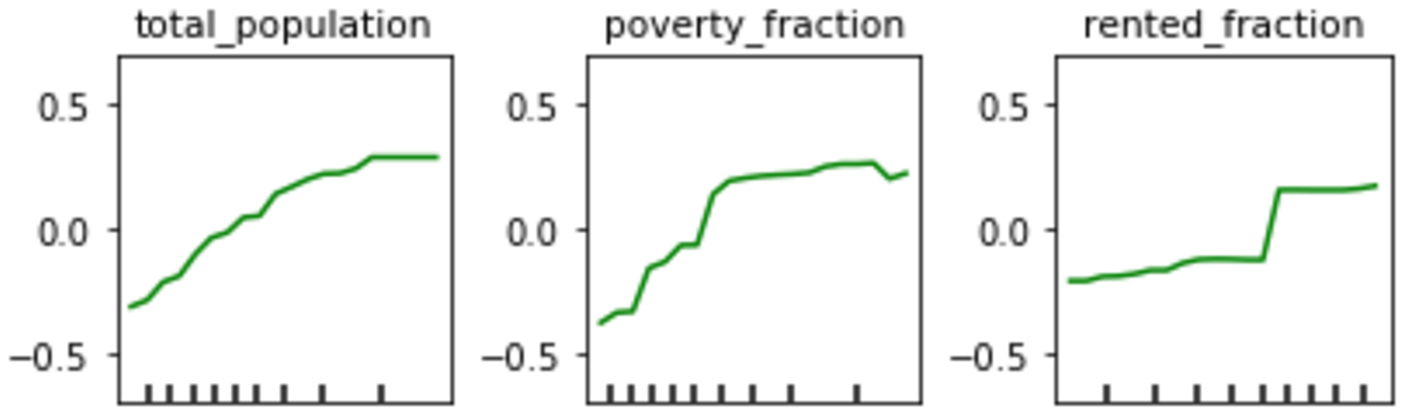, width=0.4\columnwidth}
\hskip 100pt
\psfig{file=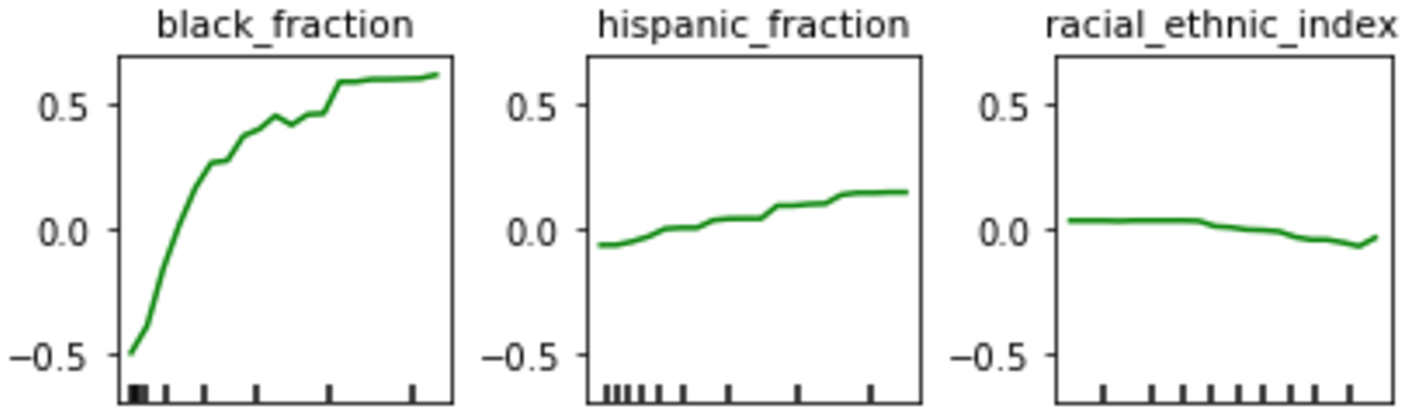, width=0.4\columnwidth}
\hskip 100pt
\psfig{file=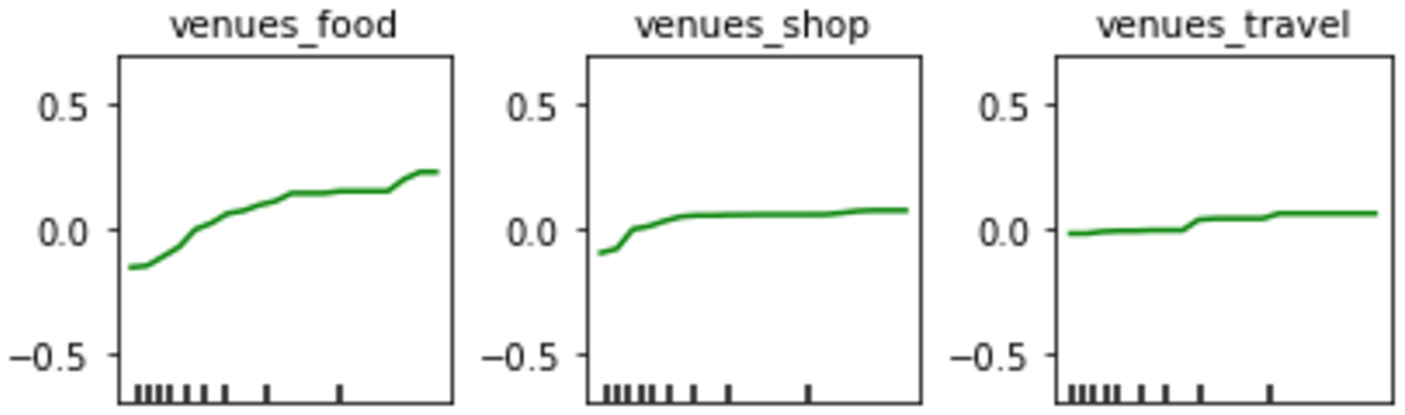, width=0.4\columnwidth}
\hskip 100pt
\psfig{file=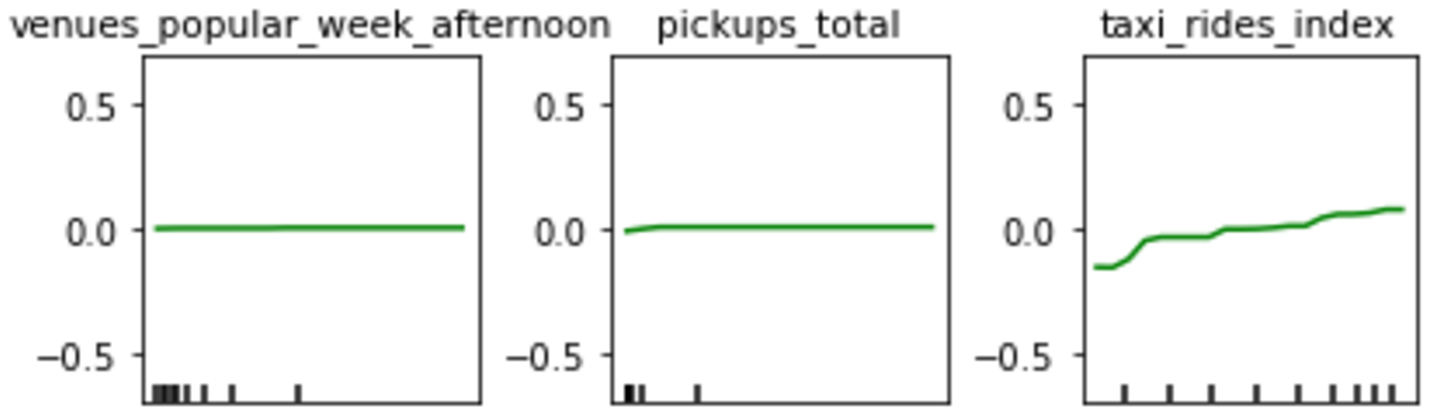, width=0.4\columnwidth}

\caption{Partial dependence plots as returned by the Gradient Boosting regressors. From top to bottom: 2015 total incidents, 2015 grand larcenies, and 2015 assaults.}
\label{fig:pdp}
\vskip -6pt
\end{figure}

Partial dependence plots visualize the marginal effect of a given single feature on the crime outcome. Figure \ref{fig:pdp} depicts the contributions of some of the features identified in the previous section as having higher predictive importance. We look at the same three types of crime: total incidents, grand larcenies, and assaults. The tick marks on the x-axis represent the deciles of the feature values in the training data. We notice that census tracts with higher population numbers, higher poverty, and higher percentage of rented houses tend to have higher crime levels. Also, neighborhoods in NYC with a higher percentage of minorities tend to have higher crime levels, with a stronger effect noticed in the assaults category. On the other hand, we also notice that highly diverse neighborhoods might be slightly safer. The POIs features exhibit strong marginal effects: especially census tracts with shopping establishments tend to experience more grand larcenies, and census tracts with food establishments tend to experience more assaults. From the spatio-temporal features, taxi drives diversity exhibits a positive relationship with the crime level across all three categories. Finally, neighborhoods with more popular venues during working day afternoons are associated with higher number of larcenies.

\subsubsection{Geographical Improvement} 

\begin{figure} [h!]
\centering
\psfig{file=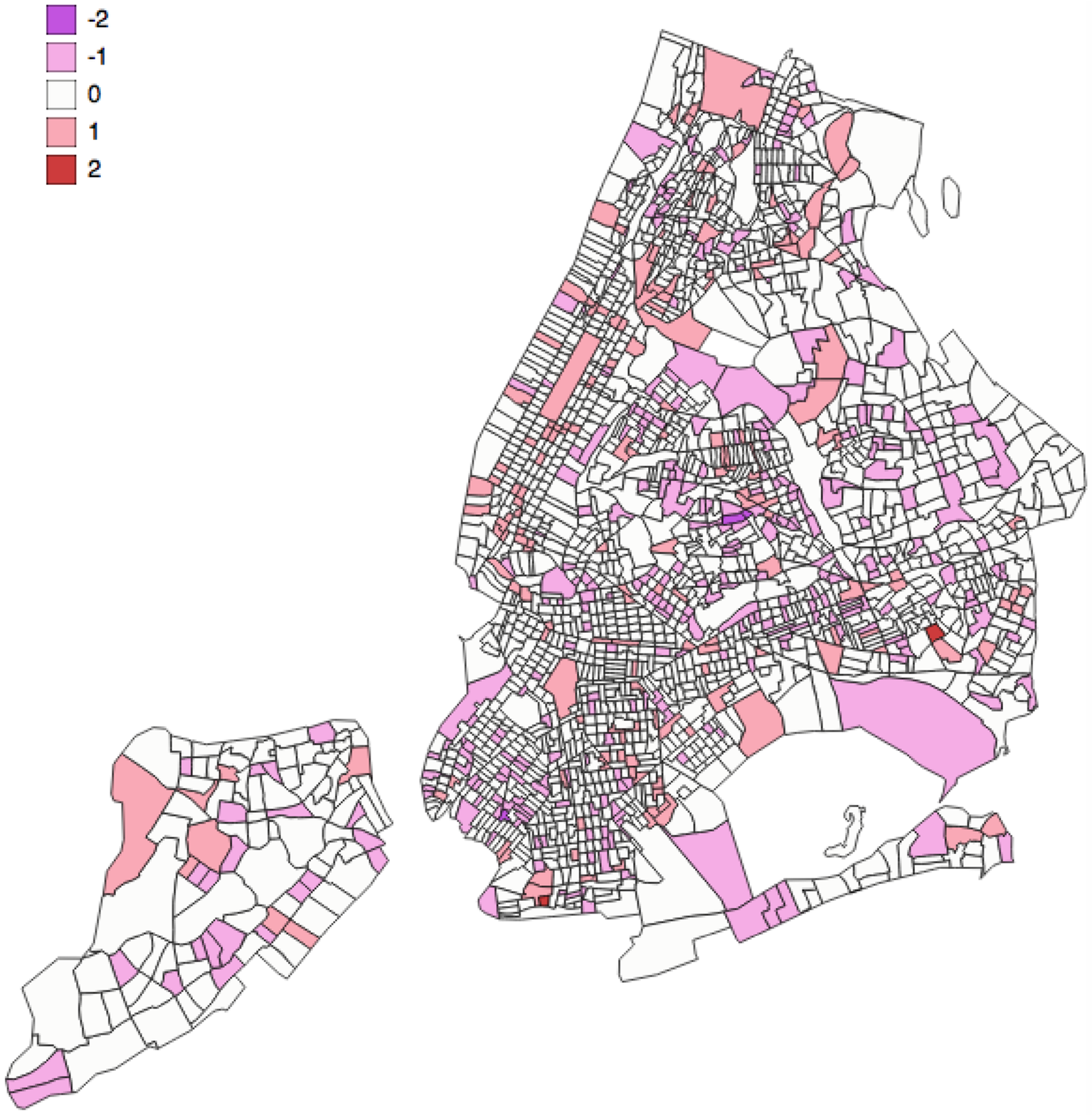, width=0.25\columnwidth}
\hskip 75pt
\psfig{file=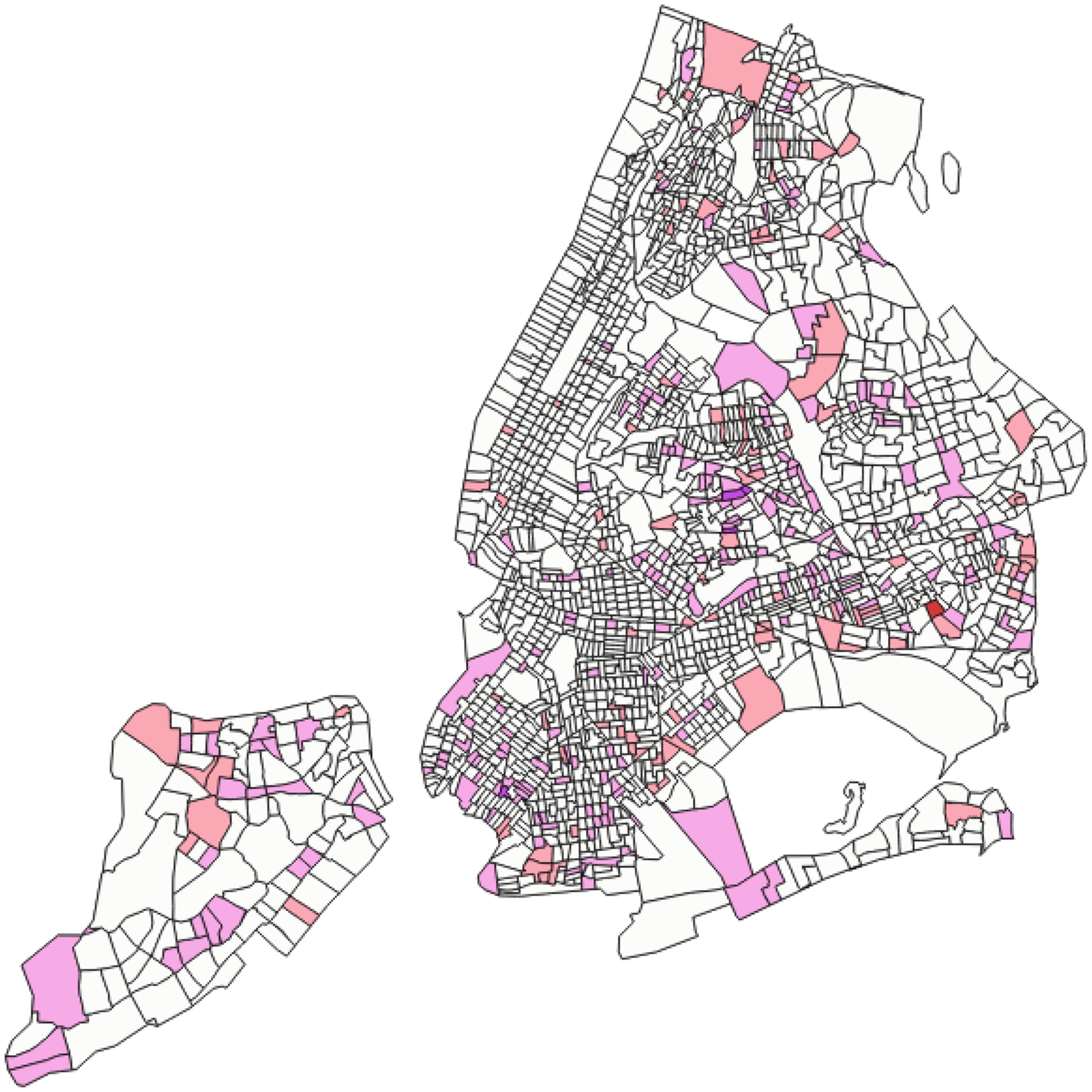, width=0.25\columnwidth}
\psfig{file=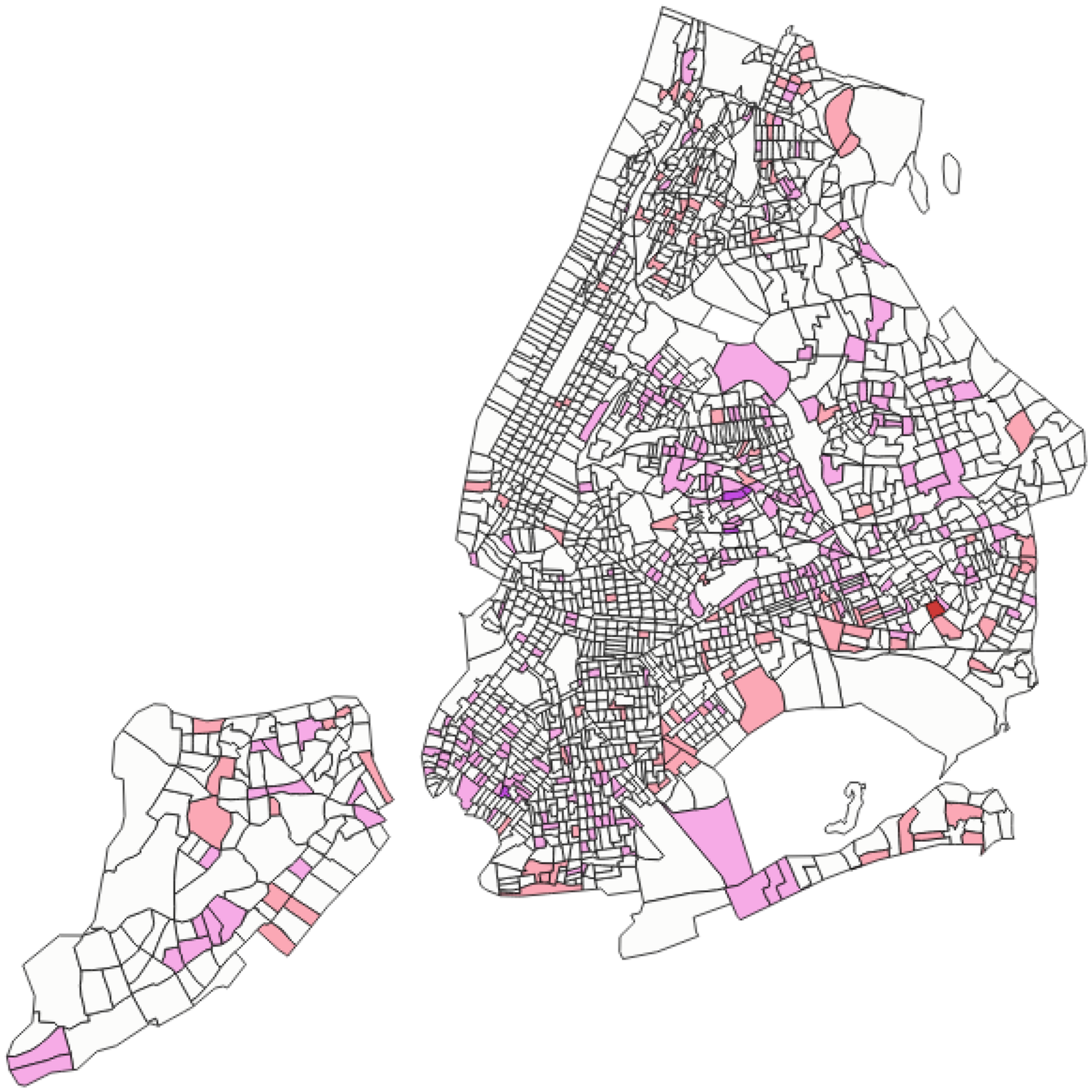, width=0.25\columnwidth}
\hskip 75pt
\psfig{file=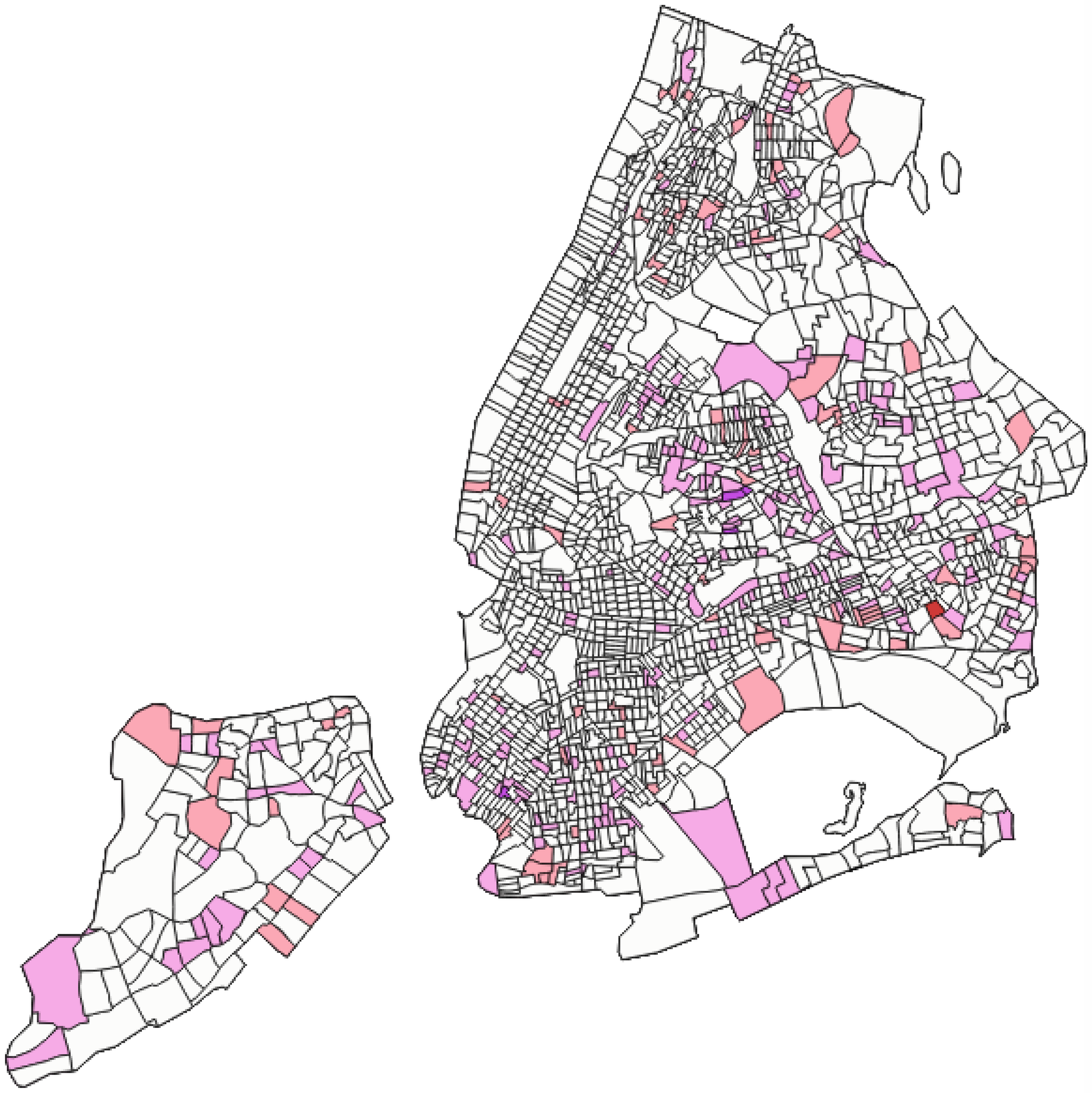, width=0.25\columnwidth}
\caption{Absolute error of predicted vs actual values for the 2015 larcenies counts per census tract. From left to right, and top to down: census (weak baseline), census + POI (strong baseline), FS + subway + taxi (human dynamics), and census + FS + subway + taxi (full model). }
\label{fig:larcenies_delta}
\vskip -6pt
\end{figure}

Finally, to understand the additive predictive power of the human dynamics features in the case of the temporal prediction, we do a deeper analysis of the residuals. Figure \ref{fig:larcenies_delta} presents the absolute error (computed as $y_i-\hat{y_i}$, rounded to integer precision) of the best models (Random Forest regressors) on the different model specifications for the 2015 grand larcenies crime category. There are 1652 (out of 2154) census tracts with an absolute error between $-0.5$ and $0.5$ in the census weak-baseline. This number increases to 1838 in the census + POI strong-baseline, and to 1850 in the full model specification. Notably, the human dynamics specification achieves a competitive high number of 1808 census tracts with low errors. The supplementary material depicts the absolute errors achieved by the remaining model specifications.

Looking at the different boroughs, the models incorporating features from FS and taxi trips consistently perform better in comparison to the census baseline in the Manhattan and Bronx boroughs, while some areas in Queens remain poorly predicted across all models. Looking at the function of the neighborhoods, these models bring improvements for parks (e.g. Central Park or Prospect Park), entertainment areas (e.g. around the NY Aquarium or the College Point Multiplex Cinemas) or the JFK airport. Between the hard-baseline incorporating only FS venues information and the model incorporating also FS check-in information, we notice improvements for instance in the Brooklyn promenade recreational areas or in the shopping areas south-east of College Point.

\section{Conclusions}
\label{sec:conclusions}

\subsection{Implications}
In this paper, long term crime prediction has been investigated at a fine-grained level, with yearly crime data being analyzed at census track level and across several crime categories. In constructing the prediction features, we exploited census data, Foursquare venues data, subway usage data, and taxi usage data by operationalizing different concepts from criminological and urban theories. Our work has both theoretical and practical implications.

First, we have identified new crime predictors derived from massive ubiquitous data sources and so extended the empirical literature in urban computing and computational social science. Our results show that, enriching the traditional census features describing the characteristics of the residential population with spatial and spatio-temporal features describing the activities of the ambient population, substantially improves the quality of the prediction models. Factors describing criminogenic places (crime attractors and crime generators) \cite{Brantingham1995} prove therefore essential for competitive crime prediction models. The highest improvement they bring has been observed in predicting crime in busy public parts of the city: recreational area and parks, shopping areas, entertainment areas, and airports. The human dynamics features improve the baseline models for the total number of incidents, for grand larcenies, and for robberies. In terms of the analyzed sources of timestamped geo-referenced human activity data, LBSNs achieve the highest predictive power. Enhancing the models with subway or with taxi data yields similar results, with the exception of the grand larcenies category, where the taxi features exhibit a higher predictive ability.  

In general, the best performing novel features for all crime incidents have been: the total number of shopping/eating/travel venues and checkins as proxy for the general popularity of that area, the number of popular venues in a normal afternoon as proxy for the temporal break-down of human activity in the area, the total number of taxi pick-ups as proxy for the population outer flow to more remote areas, and the taxi drives index as proxy for the entropy of the human movement in the area. Many of these top features can be mapped as crime attractors or crime generators and have been informed by the theories that the place and time where the offenders and victims meet are strong crime predictions \cite{Brantingham1995,Cohen1979}. While the mixed land use concept theorized by Jacobs and Newman have not been found as particularly discriminative for crime prediction in comparison to the other features, Jacob's metrics of raw human density and activity have been found to strongly improve the models. Furthermore, specific novel predictors emerge for specific crime types. 

From the census features, the metrics of concentrated disadvantage have scored highest across all crime types, which is aligned with the findings within the frameworks of the Social Disorganization Theory \cite{Pratt2005, Sampson1997}. 

On the practical side, a direct application of our results would be to have a first estimation of the safety of new developments and public spaces, for instance shopping and recreational areas. So far, crime prevention through environmental design (CPTED) \cite{Newman1996} has concentrated mostly on the attributes of the built environment (e.g. lightning, visibility, access and height of buildings) and less so on the the human activity that will be generated within the new created space. A derived product can also be used by individuals (either locals or tourists) to assess the incidents risk when traveling, going out or going shopping to new areas that they are not familiar with. 
Furthermore, an extension of the presented prediction models could be operationally deployed by local police agencies for short term risk assessment and effective deployment of patrol resources. Forces on the ground could better target specific types of crimes expected in a small geographical area. Current software solutions like PredPol\footnote{\url{http://www.predpol.com}} only work effectively for burglaries and rely mostly on recent crime (near-repeat victimization) and less on attributes of the environment or of the ambient population. Our findings therefore expand the scope to street crimes and utilize further information on the time and place of potential crimes.

\subsection{Discussion}
Our results add to existing body of empirical literature. Compared to \cite{Traunmueller2014}, we go beyond correlation analysis between human dynamics features and crime counts, and explore a highly multi-variate non-linear prediction setup. While our diversity and ratio metrics do not match one-to-one, similar metrics to the ones used in this work make it also to our top most discriminative features, e.g. the age diversity index. Yet, we are careful to interpret the results as supporting or opposing Jacobs'/Newman's theories, as the relationships between the population density and diversity and crime are non-causal and non-linear in our case. 
Similarly to \cite{Wang2016}, we generally find that features derived from the venues consistently improve the basis models based solely on census data. In comparison to their work, we go beyond simple POI counts and derive second-order features from Foursquare informed by works in criminology and urban computing, and also additionally exploit sources of mobility patterns: subway and taxi drives. While they employed standard regression models, we employed non-parametric machine learning models, which boosted the performance. 
Also, similar to \cite{Bogomolov2014}, we demonstrate the potential of human dynamics features for the crime domain. In comparison to their work, we leverage Foursquare, subway, and taxi data instead of telecommunication data, which is arguably easier to access for research and poses less ethical questions. We also run a more comprehensive analysis leveraging: (1) more extensive datasets in terms of temporal coverage of the collected data (weeks versus years) and (2) several machine learning techniques for a more difficult prediction task (regression versus binary classification). Finally, compared to all of these previous works, we are the only ones to take deeper dives into the different crime types and do careful model interpretation.  

We also contribute to the methodological literature. The main strengths of the employed machine learning algorithms are their very high predictive power and their ability to deal with heterogeneous data sources and potentially collinear factors. This opens the door for future incorporation of new variables as features for which, a priori, there is no substantive theory underlying their association with crime, but might be found to have a strong predictive power. The model interpretation techniques available for tree-based ensemble models (feature importance rankings, partial dependence plots) make the models more transparent and offer insights in terms of the predictive power of each feature. On the weaknesses side, as for any supervised learning technique, the presented models can be used for prediction, but not for inferring a causal effect between the features and the dependent variable.

We should acknowledge the geographical (more urban areas) \cite{Hecht2014} and social bias (younger, more educated, wealthier users) \cite{Cranshaw2012} of Foursquare in general, though the choice of NYC (as the city with most activity on Foursquare) and of the complete aggregated information on venues level (as opposed to incomplete extracts of checkins on users level which are common in literature) are good mitigation approaches. Quantifying such biases would become relevant once comparing different locations \cite{Quattrone2015}, but are for now out of scope for this study. 

Also, we ought to acknowledge the reporting bias present in the crime data itself. Bias in police records can be attributed to: (1) levels of community trust in police, in case of self-reported crimes, and (2) patrolling focus on certain ethnic groups and neighborhoods, in case of police-reported crimes. Even if we do not have the ambition of solving the perpetuation of racial biases in police work, we should note that this can introduce dangerous biases \cite{Lum2016}. Training models on biased historical data and having police focus on certain communities, will lead to even more arrests of minorities, but will not lead to solving the crime problem. The solution is not trivial, as it lies at the heart of the interaction between the police and the communities. At higher levels of aggregation, "ground truth" crime data could be estimated from crime victimization surveys and demographically representative synthetic populations \cite{Lum2016}.

Finally, to be aligned with previous work in criminology \cite{Pratt2005} and to be able to benchmark against prior work on crime prediction \cite{Wang2016}, we have used the race of the inhabitants when crafting several of the census features for the prediction problem. A potential mitigation would be to show how well the models do without taking race into consideration, especially if planned to be used operationally. In this work, we have already shown that, for certain types of crime, models using only human mobility data can out-perform the models based only on the census data. We believe this to be a significant contribution and an important step towards more fairness in crime prediction. 

\subsection{Future Work}
For future work and to make more general claims about the predictive power of such factors for long-term crime prediction globally, we plan to apply the same methodology on data from other major cities around the globe. Furthermore, the models can be enhanced by exploiting further ubiquitous data sources describing the pulse of our cities, like additional social media signals, 311 calls, and IoT devices. Especially for some specific type of crimes, like burglaries and vehicles thefts, incorporating spatial features describing the built environment (houses, streets, land use, etc.), has the potential to improve the models significantly. Finally, introducing temporal crime correlates (weather data, near-repeat patterns, entertainment events, etc.) has support in criminology and the potential to improve our prediction models towards short-term prediction.

\section{Abbreviations}
ACS - American Community Survey\\
BMT - Brooklyn-Manhattan Transit Company\\
ET - Extra-Trees\\
GB - Gradient Boosting\\
IND - Independent Subway System\\
IRT - Interborough Rapid Transit Company\\
LBSN - Location-Based Social Networks\\
MTA - Metropolitan Transportation Authority\\
MSE - Mean Squared Error \\
NYC - New York City\\
NYPD - New York Police Department\\
RF - Random Forests\\

\begin{backmatter}

\section*{Authors' information}
CK is a doctoral researcher at the Department of Management, Technology, and Economics (D-MTEC) of the Swiss Federal Institute of Technology in Zurich (ETH Zurich) and has been awarded the IBM PhD Fellowship for her academic achievements. She holds a MSc with Honors from the Elite Graduate Program in Software Engineering, a joint initiative of three Bavarian universities: Technical University of Munich, Ludwig Maximilian University of Munich, and University of Augsburg. 

IP was a post-doctoral researcher at the Department of Management, Technology, and Economics (D-MTEC) of the Swiss Federal Institute of Technology in Zurich (ETH Zurich). She holds a PhD in Management Science from the same institution, and a MSc in Electrical Engineering and Computer Science from Ss. Cyril and Methodius University Skopje. 

\section*{Author's contributions}
CK collected part of the data, designed the experiments, carried out the analysis, prepared the figures, and wrote the manuscript. IP collected part of the data and discussed with CK the initial design of the study.
Both authors read and approved the final manuscript.

\section*{Competing interests}
The authors declare that they have no competing interests.

\section*{Funding}
Not applicable.

\section*{Acknowledgements}
We would like to thank Raquel Ros{\'e}s Br{\"u}ngger for her valuable insights into criminology. 

\section*{Availability of data and material}
All data (crime dependent variable and processed features) can be requested directly from the authors.


\bibliographystyle{bmc-mathphys} 
\bibliography{epj}      




\end{backmatter}
\end{document}